\documentclass[sigconf,nonacm]{aamas} 

\usepackage{balance} 

\usepackage[T1]{fontenc}
\usepackage{amsmath}
\usepackage{centernot}
\usepackage{xspace}
\usepackage{multirow}
\usepackage[normalem]{ulem}
\usepackage{lineno}
\usepackage{comment}

\usepackage[UKenglish]{babel}
\usepackage{multicol}
\usepackage{tikz}
\usetikzlibrary{arrows,decorations,shapes,automata,positioning,decorations.pathmorphing,calc,patterns,backgrounds}
\colorlet{colorfilltopicclass}{yellow}
\tikzset{
  every picture/.style = {
    thick,
    ->,
    >=stealth',
  }
  ,
  within/.style = {
    fill = white,
    inner sep = 2pt
  }
  ,
  topicclass/.style = {
    -,
    draw=colorfilltopicclass!75!black,
    fill=colorfilltopicclass,
    fill opacity=0.2
  }
  ,
  cross line/.style={
    preaction = {
      draw=white,
      -,
      line width=4pt}
  }
  ,
  frame rectangle/.style = {
    framed,
    draw=black,
    rounded corners,
  }
  ,
  myworld/.style = {
    rectangle,
    rounded corners = 5,
    draw = black!100,
    fill = lightgray!40,
    inner sep = 4pt,
    minimum size = 12pt,
    font = \footnotesize,
  }
  ,
  myworld-label/.style = {
    font = \scriptsize,
    outer sep = 1pt,
    inner sep = 1pt
  }
  ,
  myarrow/.style = {
    black,
    solid
  }
  ,
  myarrow-label/.style = {
    font = \scriptsize
  }
}

\usepackage{hyperref}

\newcommand{\titulo}{(Arbitrary) Partial Communication}

\hypersetup{
  pdfstartview      = {FitH},
  pdfpagelayout     = {OneColumn},
  bookmarksnumbered = {true},
  naturalnames      = {true},
  hypertexnames     = {false},
}

\addto\extrasUKenglish{%

}

\newcommand{\autorefp}[1]{(\autoref{#1})}

\usepackage{enumitem}

\newcommand{\itmformat}{\bfseries \itshape}

\setlist[enumerate,1]{label={\itmformat (\roman*)}}
\setlist[enumerate,2]{label*={\itmformat \arabic*}}

\newlist{inlineenum}{enumerate*}{1}
\setlist[inlineenum,1]{label={\itmformat (\roman*)}}

\newlist{inlineenumco}{enumerate*}{1}
\setlist[inlineenumco,1]{label={\itmformat (\arabic*)}}

\newlist{compactenumerate}{enumerate}{1}
\setlist[compactenumerate,1]{label={\itmformat (\roman*)}, leftmargin=1.5em, rightmargin=0em, topsep=0.25em, itemsep=0.25em}

\setlist[itemize,1]{label=\textbullet}
\setlist[itemize,2]{label=--}

\newlist{compactitemize}{itemize}{3}
\setlist[compactitemize,1]{label=\textbullet, leftmargin=1.5em, rightmargin=0em, topsep=0.25em, itemsep=0.25em}
\setlist[compactitemize,2]{label=--, leftmargin=1.25em, rightmargin=0em, topsep=0.25em, itemsep=0.25em}
\setlist[compactitemize,3]{label=$\cdot$, leftmargin=0.25em, rightmargin=0em, topsep=0.25em, itemsep=0.25em}

\newlist{itemizecom}{enumerate}{1}
\setlist[itemizecom,1]{label=--, leftmargin=1em, rightmargin=0em, topsep=0.25em, itemsep=0.25em, before*=\small}

\usepackage{algorithm}
\usepackage[noend]{algpseudocode}

\algtext*{EndIf}
\algtext*{EndFor}

\algnewcommand\algorithmiccase{\tb{case}}
\algdef{SE}[CASE]{Case}{EndCase}[1]{\algorithmiccase\ #1}{\algorithmicend\ \algorithmiccase}%
\algtext*{EndCase}

\makeatletter
\newenvironment{breakablealgorithm}
  {
   \begin{center}
     \refstepcounter{algorithm}
     \hrule height.8pt depth0pt \kern2pt
     \renewcommand{\caption}[2][\relax]{
       {\raggedright\tb{\fname@algorithm~\thealgorithm} ##2\par}%
       \ifx\relax##1\relax 
         \addcontentsline{loa}{algorithm}{\protect\numberline{\thealgorithm}##2}%
       \else 
         \addcontentsline{loa}{algorithm}{\protect\numberline{\thealgorithm}##1}%
       \fi
       \kern2pt\hrule\kern2pt
     }
  }{
     \kern2pt\hrule\relax
   \end{center}
  }
\makeatother

\colorlet{txthlcolour}{orange!25!yellow}

\title{\titulo}

\author{Rustam Galimullin}
\affiliation{%
  \institution{Department of Information Science and Media Studies, Universitetet i Bergen}%
  \city{Bergen}%
  \country{Norway}%
}
\email{Rustam.Galimullin@uib.no}

\author{Fernando R. Vel{\'a}zquez-Quesada}
\affiliation{%
  \institution{Department of Information Science and Media Studies, Universitetet i Bergen}%
  \city{Bergen}%
  \country{Norway}%
}
\email{Fernando.VelazquezQuesada@uib.no}

\titlenote{This is a slightly extended version of the same title paper that will appear in AAMAS 2023. This version contains a small appendix with proofs that, for space reasons, do not appear in the AAMAS 2023 version.}

\begin{abstract}
  Communication within groups of agents has been lately the focus of research in dynamic epistemic logic (DEL). This paper studies a recently introduced form of \emph{partial} (more precisely, \emph{topic-based}) \emph{communication}. This type of communication allows for modelling scenarios of multi-agent collaboration and negotiation, and it is particularly well-suited for situations in which sharing all information is not feasible/advisable. After presenting results on invariance and complexity of model checking, the paper compares partial communication to public announcements, probably the most well-known type of communication in DEL. It is shown that the settings are, update-wise, incomparable: there are scenarios in which the effect of a public announcement cannot be replicated by partial communication, and vice versa. Then, the paper shifts its attention to \emph{strategic} topic-based communication. It does so by extending the language with a modality that quantifies over the topics the agents can `talk about'. For this new framework, it provides a complete axiomatisation, showing also that the new language's model checking problem is \textit{PSPACE}-complete. The paper closes showing that, in terms of expressivity, this new language of arbitrary partial communication is incomparable to that of arbitrary public announcements.
\end{abstract}

\ccsdesc[500]{Theory of computation~Modal and temporal logics}

\keywords{partial communication, arbitrary partial communication, distributed knowledge, public announcement, dynamic epistemic logic, epistemic logic}

\renewcommand{\arraystretch}{1.3} 

\renewcommand{\emptyset}{\varnothing}

\newcommand{\ti}[1]{\textit{#1}}
\newcommand{\tb}[1]{\textbf{#1}}

\newenvironment{ltabular}[1]
{\begin{flushleft}\begin{tabular}{#1}}
{\end{tabular}\end{flushleft}}

\newenvironment{lbtabular}[1]
{\begin{flushleft}\begin{tabular}[b]{#1}}
{\end{tabular} \end{flushleft}}

\newenvironment{ctabular}[1]
{\begin{center}\begin{tabular}{#1}}
{\end{tabular} \end{center}}

\newenvironment{smallctabular}[1]
{\begin{center}\begin{small}\begin{tabular}{#1}}
{\end{tabular}\end{small}\end{center}}

\newcommand{\msparagraph}[1]{\medskip\noindent\tb{#1}\,}
\newcommand{\ssparagraph}[1]{\smallskip\noindent\tb{#1}\,}

\newcommand{\nsparagraph}[1]{\noindent\tb{#1}\,}

\newcommand{\tupla}[1]{\left\langle #1 \right\rangle}
\newcommand{\tuplan}[1]{\langle #1 \rangle}
\newcommand{\set}[1]{\left\{ #1 \right\}}
\newcommand{\setn}[1]{\mathopen{\{} {#1} \mathclose{\}}}
\newcommand{\card}[1]{|#1|}

\newcommand{\power}[1]{\wp({#1})}
\newcommand{\LA}{\ensuremath{\mathcal{L}}\xspace}
\newcommand{\limp}{\rightarrow}
\newcommand{\ldimp}{\leftrightarrow}
\newcommand{\lldimp}{\,\ldimp\,}
\newcommand{\sub}{\operatorname{sub}}
\newcommand{\truthset}[2]{\left\llbracket #2 \right\rrbracket^{#1}}

\newcommand{\Nat}{\mathbb{N}}

\newcommand{\mDia}[1]{\mathop{\langle#1\rangle}}
\newcommand{\mBox}[1]{\mathop{[#1]}}
\newcommand{\mmBox}[2]{\mBox{#1}#2}
\newcommand{\mmDia}[2]{\mDia{#1}#2}

\newcommand{\ssi}{\text{iff}}

\newcommand{\ssidef}{\ensuremath{\mathrel\ssi_{\mathit{def}}}}
\newcommand{\qssidefq}{\quad\ssidef\quad}

\newcommand{\qtnq}[1]{\quad\text{#1}\quad}

\newcommand{\tr}{\mathit{tr}}               
\newcommand{\bisimsymbol}{\rightleftarrows} 
\newcommand{\cobisim}{\bisimsymbol_C}       

\newcommand{\mc}{\mathit{MC}}             

\newcommand{\pa}{\ensuremath{\mathtt{P}}\xspace}
\newcommand{\spa}{\ensuremath{\mathtt{Q}}\xspace}

\newcommand{\formatagent}[1]{\mathtt{#1}}

\newcommand{\ag}{\ensuremath{\formatagent{A}}\xspace}
\newcommand{\agi}{\formatagent{i}}
\newcommand{\agj}{\formatagent{j}}
\newcommand{\agk}{\formatagent{k}}
\newcommand{\aga}{\formatagent{a}}
\newcommand{\agb}{\formatagent{b}}
\newcommand{\agc}{\formatagent{c}}
\newcommand{\sag}{\ensuremath{\formatagent{G}}}
\newcommand{\sen}{\ensuremath{\formatagent{S}}}

\newcommand{\opK}{\operatorname{K}}
\newcommand{\opgDist}{\operatorname{D}}

\newcommand{\dom}[1]{\mathfrak{D}(#1)}
\newcommand{\R}{R}
\newcommand{\Rp}{R'}

\newcommand{\sbi}[1]{{#1}_{\agi}}

\newcommand{\sbk}[1]{{#1}_{\agk}}
\newcommand{\sba}[1]{{#1}_{\aga}}

\newcommand{\sbDs}[2]{{#1}_{#2}}
\newcommand{\sbDg}[1]{{#1}_{\sag}}

\newcommand{\modK}[1]{\mathop{\opK_{#1}}}

\newcommand{\mmKi}[1]{\modK{\agi}#1}

\newcommand{\mmKa}[1]{\modK{\aga}#1}
\newcommand{\mmKb}[1]{\modK{\agb}#1}

\newcommand{\modKD}[1]{\mathop{\widehat{\opK}_{#1}}}

\newcommand{\mmKDa}[1]{\modKD{\aga}#1}

\newcommand{\modDs}[1]{\mathop{\opgDist_{#1}}}
\newcommand{\modDg}{\modDs{\sag}}
\newcommand{\mmDg}[1]{\modDg#1}
\newcommand{\mmDs}[2]{\modDs{#1}#2}

\newcommand{\modDsr}[2]{\mathop{\opgDist^{#2}_{#1}}}
\newcommand{\mmDsr}[3]{\modDsr{#1}{#2}#3}

\newcommand{\knonfu}[2]{\sim^{#1}_{#2}}

\newcommand{\LAd}{\ensuremath{{\LA}}\xspace}
\newcommand{\LAde}[1]{\ensuremath{\LA_{#1}}\xspace}
\newcommand{\LAdSSE}{\LAde{\symbSSE{\sen}{\!\chi}}}
\newcommand{\LAdPA}{\LAde{\symbPA{\xi}}}
\newcommand{\LAdSAE}[1]{\ensuremath{{\LA^{\ast#1}_{\symbSSE{\sen}{\!\chi}}}}\xspace}
\newcommand{\LAdAPA}{\ensuremath{{\LA^{\ast}_{\symbPA{\xi}}}}\xspace}

\newcommand{\ax}[3]{\textnormal{\textsf{#1}}$^{#2}_{#3}$\xspace}

\newcommand{\LO}{\ensuremath{\mathsf{L}}\xspace}

\newcommand{\dsrMP}{\ax{MP}{}{}}

\newcommand{\LOd}{\ensuremath{\LO}\xspace}
\newcommand{\LOde}[1]{\ensuremath{\LO_{#1}}\xspace}

\newcommand{\LOdSSE}{\LOde{\symbSSE{\sen}{\!\chi}}}
\newcommand{\dsaSSE}[1]{\ax{A}{#1}{\symbSSE{\sen}{\!\chi}}}
\newcommand{\dsreSSE}{\ax{RE}{}{\symbSSE{\sen}{\!\chi}}}

\newcommand{\symbSEE}[1]{{#1}!}
\newcommand{\relSEE}[2]{#1^{\symbSEE{#2}}}

\newcommand{\symbSSE}[2]{{#1{:}\,#2}!}

\newcommand{\mopSSE}[3]{{#1_{\symbSSE{#2}{#3}}}}
\newcommand{\relSSE}[3]{#1^{\symbSSE{#2}{#3}}}
\newcommand{\modboxSSE}[2]{\mmBox{\symbSSE{#1}{#2}}{}}
\newcommand{\moddiaSSE}[2]{\mmDia{\symbSSE{#1}{#2}}{}}
\newcommand{\mmSSE}[3]{\modboxSSE{#1}{#2}#3}
\newcommand{\mmdiaSSE}[3]{\moddiaSSE{#1}{#2}#3}

\newcommand{\symbPA}[1]{#1!}
\newcommand{\mopPA}[2]{#1_{\symbPA{#2}}}
\newcommand{\relPA}[2]{#1^{\symbPA{#2}}}
\newcommand{\modboxPA}[1]{\mmBox{\symbPA{#1}}{}}
\newcommand{\moddiaPA}[1]{\mmDia{\symbPA{#1}}{}}
\newcommand{\mmPA}[2]{\modboxPA{#1}#2}
\newcommand{\mmdiaPA}[2]{\moddiaPA{#1}#2}

\newcommand{\symbSAE}[1]{{#1{:}\,\ast}!}

\newcommand{\modboxSAE}[1]{\mmBox{\symbSAE{#1}}{}}
\newcommand{\moddiaSAE}[1]{\mmDia{\symbSAE{#1}}{}}
\newcommand{\mmSAE}[2]{\modboxSAE{#1}#2}
\newcommand{\mmdiaSAE}[2]{\moddiaSAE{#1}#2}

\newcommand{\replace}[2]{#2}

\begin{document}

\newtheorem{fact}[theorem]{Fact}

\pagestyle{fancy}
\fancyhead{}

\maketitle

\section{Introduction}\label{sec:intro}

Epistemic logic (\ti{EL}; \cite{Hintikka1962}) is a powerful framework for representing the individual and collective knowledge/beliefs of a group of agents. When using relational `Kripke' models, its crucial idea is the use of uncertainty for defining knowledge. Indeed, such structures assign to each agent a binary relation indicating \emph{indistinguishability} among epistemic possibilities. Then, it is said that agent $\agi$ knows that $\varphi$ is the case (syntactically: $\mmKi{\varphi}$) when $\varphi$ holds in all situations $\agi$ considers possible. Despite its simplicity, \ti{EL} has become a widespread tool, contributing to the formal study of complex multi-agent epistemic phenomena in philosophy \cite{PhilStu:EpisLog}, computer science \cite{FaginHalpernMosesVardi1995}, AI \cite{MeyervanDerHoek1995elaics} and economics \cite{deBruin2010}.

One of the most appealing aspects of \ti{EL} is that it can be used for reasoning about information change. This has been the main subject of dynamic epistemic logic (\ti{DEL}; \cite{vanDitmarschEtAl2007,vanBenthem2011ldii}), a field whose main feature is that actions are semantically represented as operations that transform the underlying semantic model. Within \ti{DEL}, one of the simplest meaningful epistemic actions is that of a \emph{public announcement}: an external source providing the agents with truthful information in a fully public way \cite{Plaza1989,GerbrandyGroeneveld1997}. Yet, the agents do not need \replace{to wait for some}{an} external entity to feed them with facts: they can also share their individual information with one another. This is arguably a more suitable way of modelling information change in multi-agent (and, in particular, distributed) systems. Agents might occasionally receive information `from the outside', but the most common form of interaction is the one in which they themselves engage in `conversations' for sharing what they have come to know so far. It is this form of information exchange that allows independent entities to engage in collaboration, negotiation and so on.

Communication between agents can take several forms, \replace{and some of these variations have been}{with some of these alternatives} explored within \replace{the \ti{DEL} framework}{\ti{DEL}}. A single agent might share all her information with everybody, as modelled in \cite{Baltag2010slides}. Alternatively, a group of agents might share all their information only among themselves, as represented by the action of ``resolving distributed knowledge'' \replace{studied in}{from} \cite{AgotnesWang2017}. One can even think about this form of communication not as a form of `sharing', but rather as a form of `taking' \cite{BaltagSmets2020,BaltagSmets2021}, which allows the study of public and private forms of reading someone else's information (e.g., hacking). 

All these approaches for inter-agent communication have a common feature: \replace{when sharing, the}{the sharing} agents share \emph{all their information}. This is of course useful, as then one can reason about the best the agents can do together. But there are also scenarios (\replace{more common, one can argue}{arguably more common}) in which sharing all her available information might not be feasible or advisable for an agent. For the first, there might be constraints on the communication channels; for the second, agents might not be in a cooperative scenario, but rather in a competitive one. In such cases, one would be rather interested in studying forms of \emph{partial} communication, through which agents share only `part of what they know'. There might be different ways to make precise what each agent shares, but a natural one is to assume that the `conversation' is relative to a subject/topic, defined by a given formula $\chi$. Introduced in \cite{Velazquez2022}, this type of communication allows \replace{for}{} a more realistic modelling of scenarios of multi-agent collaboration and negotiation. The first part of this paper studies computational aspects of this \emph{partial communication} framework. It starts \autorefp{sec:background} by recalling the main definitions and axiom system, providing then novel invariance and model checking results. After that, it discusses \autorefp{sec:SSE.vs.PAL} the setting's relationship with the public announcement framework, showing that although the languages are equally expressive, \replace{there are cases in which}{in general} the operations cannot mimic each other.

Still, in truly competitive scenarios, what matters the most is the decision of \emph{what} to share. In other words, what matters is \replace{being able}{} to reason about \emph{strategic} topic-based communication. \replace{In order to}{To} do so, the second part of this paper introduces a framework for quantifying over the conversation's topic. It presents \autorefp{sec:SAE} the basic definitions, providing then results on invariance, axiom system, expressivity and \replace{the complexity of its}{model checking}\replace{problem}{}. After that, it compares this new setting with that of arbitrary public announcements, proving that the languages are, expressivity-wise, incomparable. \autoref{sec:discussion} contrasts choices made with their alternatives, and \autoref{sec:end} summarises the paper's contents, discussing also further research lines.

\section{Background}\label{sec:background}

Throughout this text, let \ag be a finite non-empty group of agents, and let \pa be a non-empty enumerable set of atomic propositions.

\begin{definition}[Model]\label{def:model}
  A \emph{multi-agent relational model} (from now on, a model) is a tuple $M = \tupla{W, \R, V}$ where $W$ (also denoted as $\dom{M}$) is a non-empty set of objects called \emph{possible worlds}, $\R = \set{\sbi{\R} \subseteq W \times W \mid \agi \in \ag}$ assigns a binary \emph{``indistinguishability''} relation on $W$ to each agent in $\ag$ (for $\sag \subseteq \ag$, define $\sbDg{\R} := \bigcap_{\agk \in \sag} \sbk{\R}$), and $V:\pa \to \power{W}$ is an atomic valuation (with $V(p)$ the set of worlds in $M$ where $p \in \pa$ holds). A pair $(M, w)$ with $M$ a model and $w \in \dom{M}$ is a \emph{pointed model}, with $w$ being the \emph{evaluation point}. \replace{We call}{A} model $M$ \replace{}{is} \emph{finite} \replace{, if}{iff} both $W$ and $\bigcup_{w \in W}\{p \in \pa \mid w \in V(p)\}$ are finite.  If \replace{model}{} $M = \tupla{W, \R, V}$ is finite, \replace{then the \emph{size} of $M$, denoted by $\card{M}$,}{its \emph{size} (notation: $|M|$)} is \replace{defined as}{}
  $\card{W} + \sum_{\agi \in \ag}\card{\sbi{\R}} + \sum_{w \in W} \card{\set{p \in \pa \mid w \in V(p)}}$.
\end{definition}

In a model, the agents' indistinguishability relations are arbitrary. In particular, they need to be neither reflexive nor symmetric nor Euclidean nor transitive. Hence, \emph{``knowledge''} here is neither truthful nor positively/negatively introspective. It rather corresponds simply to ``what is true in all the agent's epistemic alternatives''.

\begin{definition}[Relative expressivity]
  Let $\LA_1$ and $\LA_2$ be two languages interpreted over pointed models. It is said that \emph{$\LA_2$ is at least as expressive as $\LA_1$} (notation: $\LA_1 \preccurlyeq \LA_2$) if and only if for every $\alpha_1 \in \LA_1$ there is $\alpha_2 \in \LA_2$ such that $\alpha_1$ and $\alpha_2$ have the same truth-value in every pointed model. Write $\LA_1 \approx \LA_2$ when $\LA_1 \preccurlyeq \LA_2$ and $\LA_2 \preccurlyeq \LA_1$; write $\LA_1 \prec \LA_2$ when $\LA_1 \preccurlyeq \LA_2$ and $\LA_2 \not\preccurlyeq \LA_1$; write $\LA_1 \asymp \LA_2$ when $\LA_1 \not\preccurlyeq \LA_2$ and $\LA_2 \not\preccurlyeq \LA_1$.
\end{definition}

Note: to show $\LA_1 \not\preccurlyeq \LA_2$, it is enough to find two pointed models that agree in all \replace{formulas in $\LA_2$ but can be distinguished by a formula in $\LA_1$}{$\alpha_1 \in \LA_2$ but can be distinguished by some $\alpha_2 \in \LA_1$}.

\subsection{Basic language}\label{sbs:LAd}

Here is this paper's basic language for describing pointed models.

\begin{definition}[Language \LAd]
\label{def:basiclang}
  Formulas $\varphi, \psi$ in \LAd are given by
  \[ \varphi, \psi ::= p \mid \lnot \varphi \mid \varphi \land \psi \mid \mmDg{\varphi} \]
  for $p \in \pa$ and $\emptyset \subset \sag \subseteq \ag$. Boolean constants and other Boolean operators are defined as usual. Define also $\mmKi{\varphi} := \mmDs{\set{\agi}}{\varphi}$. The \emph{size} of $\varphi$, denoted $\card{\varphi}$, is \replace{defined as follows:}{given by} $\card{p} := 1$, $\card{\lnot \varphi} = \card{\mmDg{\varphi}} := \card{\varphi} + 1$ and $\card{\varphi \land \psi} := \card{\varphi} + \card{\psi} + 1$.
\end{definition}

The language \LAd contains a modality $\modDg$ for each non-empty group of agents $\sag \subseteq \ag$. Formulas of the form $\mmDg{\varphi}$ are read as ``the agents in $\sag$ know $\varphi$ distributively''; thus, $\mmKi{\varphi}$ is read as \replace{``$\agi$ knows $\varphi$ distributively'', i.e., }{``agent $\agi$ knows $\varphi$''}. The language's semantic interpretation is as follows.

\begin{definition}[Semantic interpretation for \LAd]
  Let $(M, w)$ be a pointed model with $M = \tupla{W, \R, V}$. The satisfiability relation $\Vdash$ between $(M, w)$ and formulas in \LAd is defined inductively. Boolean cases are as usual; for the rest,
  \begin{ltabular}{l@{\qssidefq}l}
    $(M, w) \Vdash p$              & $w \in V(p)$, \\
    $(M, w) \Vdash \mmDg{\varphi}$ & for all $u \in W$, if $\sbDg{\R}wu$ then $(M, u) \Vdash \varphi$. \\
  \end{ltabular}
  Given a model $M$ and a formula $\varphi$,
  \begin{compactitemize}
    \item the set $\truthset{M}{\varphi} := \set{w \in \dom{M} \mid (M, w) \Vdash \varphi}$ contains the worlds in $\dom{M}$ in which $\varphi$ holds (also called $\varphi$-worlds);
    \item the (note: equivalence) relation
    \[ {\knonfu{M}{\varphi}} := (\truthset{M}{\varphi} \times \truthset{M}{\varphi}) \cup (\truthset{M}{\lnot \varphi} \times \truthset{M}{\lnot \varphi}) \]
    splits $\dom{M}$ into (up to) two equivalence classes: one containing all $\varphi$-worlds, and the other containing all $\lnot\varphi$-worlds.
  \end{compactitemize}
  A formula $\varphi$ is valid (notation: $\Vdash \varphi$) if and only if $(M, w) \Vdash \varphi$ for every $w \in \dom{M}$ of every model $M$.
\end{definition}

\nsparagraph{Axiom system.} The axiom system \LOd \autorefp{tbl:LOd} characterises the formulas in \LAd that are valid (see, e.g., \cite{HalpernM90,FaginHalpernMosesVardi1995}). Boolean operators are taken care of by \ax{PR}{}{} and \ax{MP}{}{}. For the modality $\modDg$, while rule \ax{G}{}{\opgDist} indicates that it `contains' all validities, axiom \ax{K}{}{\opgDist} indicates that it is closed under modus ponens, and axiom \ax{M}{}{\opgDist} states that it is monotone on the group of agents (if $\varphi$ is distributively known by $\sag$, then it is also distributively known by any larger group $\sag'$).

\begin{table}[t]
  \caption{Axiom system \LOd.}
  \begin{smallctabular}{l}
    \toprule
    \begin{tabular}{l@{\;}l}
      \ax{PR}{}{}:         & $\vdash \varphi$ \;\; for $\varphi$ a propositionally valid scheme \\
      \dsrMP:              & If $\vdash \varphi$ and $\vdash \varphi \limp \psi$ then $\vdash \psi$ \\
    \end{tabular}
    \\
    \midrule
    \begin{tabular}{l@{\;}l@{\qquad}l@{\;}l}
      \ax{K}{}{\opgDist}: & $\vdash \mmDs{\sag}{(\varphi \limp \psi)} \limp (\mmDs{\sag}{\varphi} \limp \mmDs{\sag}{\psi})$ &
      \ax{G}{}{\opgDist}: & If $\vdash \varphi$ then $\vdash \mmDs{\sag}{\varphi}$ \\
      \ax{M}{}{\opgDist}: & $\vdash \mmDs{\sag}{\varphi} \limp \mmDs{\sag'}{\varphi}$ \;\; for $\sag \subseteq \sag'$ \\
    \end{tabular}
    \\
    \bottomrule
  \end{smallctabular}
  \label{tbl:LOd}
\end{table}

\begin{theorem}\label{teo:LOd}
  The axiom system \LOd \autorefp{tbl:LOd} is sound and strongly complete for \LAd.
\end{theorem}

\nsparagraph{Structural equivalence.} The following notion will be useful.

\begin{definition}[Collective $\spa$-bisimulation \cite{Roelofsen2007}]
  Let $\spa \subseteq \pa$ be a set of atoms; let $M = \tuplan{W, \R, V}$ and $M' = \tuplan{W', \Rp, V'}$ be two models. A non-empty relation $Z \subseteq W \times W'$ is a \emph{collective $\spa$-bisimulation between $M$ and $M'$} if and only if every $(u,u') \in Z$ satisfies the following.
  \begin{compactitemize}
    \item \tb{Atoms}. For every $p \in \spa$: $u \in V(p)$ if and only if $u' \in V'(p)$.
    \item \tb{Forth}. For every $\sag \subseteq \ag$ and every $v \in W$: if $\sbDg{\R}uv$ then there is $v' \in W'$ such that $\sbDg{\Rp}u'v'$ and $(v,v') \in Z$.
    \item \tb{Back}. For every $\sag \subseteq \ag$ and every $v' \in W'$: if $\sbDg{\Rp}u'v'$ then there is $v \in W$ such that $\sbDg{\R}uv$ and $(v,v') \in Z$.
  \end{compactitemize}
  Write $M \cobisim^\spa M'$ iff there is a collective $\spa$-bisimulation between $M$ and $M'$. Write $(M,w) \cobisim^\spa (M',w')$ iff a witness for $M \cobisim^\spa M'$ contains the pair $(w,w')$. Remove the superindex ``$^\spa$'' when $\spa$ is the full set of atoms $\pa$. 
  Note: the relation of collective $\spa$-bisimilarity is an equivalence relation, both on models and pointed models.
\end{definition}


The language \LAd is invariant under collective bisimilarity.

\begin{theorem}[$\cobisim$ implies \LAd-equivalence]\label{thm:cobisim.LAd}
  Let $(M,w)$ and $(M',w')$ be two pointed models. If $(M,w) \cobisim^\spa (M',w')$ then, for every $\psi \in \LAd$ containing only atoms from $\spa$,
  \begin{ctabular}{c}
    $(M,w) \Vdash \psi$ \qtnq{if and only if} $(M',w') \Vdash \psi$.
  \end{ctabular}
  \begin{proof}
    \replace{Proofs}{For} showing that a form of \replace{structural}{model} equivalence implies invariance for a language\replace{ usually proceed by structural}{, one usually uses} induction on the language's formulas.\footnote{The proofs typically start by pulling out the universal quantifier over formulas, the statement becoming \emph{``for every $\varphi$, any structurally equivalent pointed models agree on $\varphi$'s truth-value''}. This yields a stronger inductive hypothesis (IH) thanks to which the proof can go through. This will be done throughout the rest of the text.} For \replace{the case of collective}{} $\pa$-bisimilarity and \LAd, see \cite{Roelofsen2007}.
  \end{proof}
\end{theorem}

\nsparagraph{Model checking} This problem for \LA is in \textit{P} \cite[Page 67]{FaginHalpernMosesVardi1995}.

\subsection{Partial (topic-based) communication}\label{sbs:LAdSSE}

Through an action of partial communication, a group of agents $\sen \subseteq \ag$ share, with everybody, all their information about a given topic $\chi$. To define it, consider first a simpler action. After agents in $\sen$ share \emph{all their information} with everybody, 
an agent $\agi$ will consider a world $u$ possible from a world $w$ if and only if she and every agent in $\sen$ considered $u$ possible from $w$ (i.e., $\agi$'s new relation $\sbi{\relSEE{\R}{\sen}}$ is the \emph{intersection} of $\sbi{\R}$ and $\sbDs{\R}{\sen}$). In other words, after full communication, at $w$ agent $\agi$ will consider $u$ possible if and only if neither her nor any agent in $\sen$ could rule out $u$ from $w$ before the action. But if agents in $\sen$ share only `their information about $\chi$' (intuitively, only what has allowed them to distinguish between $\chi$- and $\lnot\chi$-worlds), edges between worlds agreeing in $\chi$'s truth-value are not `part of the discussion'; thus, they should not be eliminated.

\begin{definition}[Partial communication \cite{Velazquez2022}]\label{def:SSEmo}
  Let $M = \tupla{W, \R, V}$ be a model; take a group of agents $\sen \subseteq \ag$ and a formula $\chi$. The model $\mopSSE{M}{\sen}{\chi} = \tuplan{W, \relSSE{\R}{\sen}{\chi}, V}$, the result of agents in $\sen$ sharing all they know about $\chi$ with everybody, is such that
  \begin{ctabular}{c}
    $\sbi{\relSSE{\R}{\sen}{\chi}} := \sbi{\R} \cap (\sbDs{\R}{\sen} \cup {\knonfu{M}{\chi}})$.
  \end{ctabular}
  Thus, $\sbDg{\relSSE{R}{\sen}{\chi}} = \bigcap_{\agi \in \sag} \sbi{\relSSE{\R}{\sen}{\chi}} = \sbDs{R}{\sag} \cap (\sbDs{R}{\sen} \cup {\knonfu{M}{\chi}}) = \sbDs{R}{\sag \cup \sen} \cup (\sbDs{R}{\sag} \cap {\knonfu{M}{\chi}})$. Additionally, $\sbi{\relSSE{\R}{\emptyset}{\chi}} = \sbi{\R}$.
\end{definition}

\begin{definition}[Modality $\modboxSSE{\sen}{\chi}$ and language \LAdSSE \cite{Velazquez2022}]\label{def:SSEmm}
  The language \LAdSSE extends \LAd with a modality $\modboxSSE{\sen}{\chi}$ for each $\sen \subseteq \ag$ and each formula $\chi$. More precisely, define first $\LAdSSE^0 = \LAd$, and then define $\LAdSSE^{i+1}$ as the result of extending $\LAdSSE^i$ with an additional modality $\modboxSSE{\sen}{\chi}$ for $\sen \subseteq \ag$ and $\chi \in \LAdSSE^i$. The language \LAdSSE is \replace{the union of all $\LAdSSE^i$ with $i \in \Nat$}{then defined as $\bigcup_{i \in \Nat} \LAdSSE^i$}. For its semantic interpretation,
  \begin{lbtabular}{l@{\qssidefq}l}
    $(M, w) \Vdash \mmSSE{\sen}{\chi}{\varphi}$ & $(\mopSSE{M}{\sen}{\chi}, w) \Vdash \varphi$.
  \end{lbtabular}
  Defining $\mmdiaSSE{\sen}{\chi}{\varphi} := \lnot \mmSSE{\sen}{\chi}{\lnot \varphi}$ implies $\Vdash \mmdiaSSE{\sen}{\chi}{\varphi} \ldimp \mmSSE{\sen}{\chi}{\varphi}$. The size of a formula $\varphi \in \LAdSSE$ is \replace{defined}{ }as in \autoref{def:basiclang}, with \replace{an}{the} additional clause $\card{\mmSSE{\sen}{\chi}{\varphi}} := \card{\chi} + \card{\varphi} + 1$.
\end{definition}

Further motivation and details on partial communication can be found in \cite{Velazquez2022}. Still, here are two revealing properties: $\Vdash \mmSSE{\sen}{\chi_1}{\varphi} \ldimp \mmSSE{\sen}{\chi_2}{\varphi}$ for $ \Vdash \chi_1 \ldimp \chi_2$ (logically equivalent topics have the same communication effect) and $\Vdash \mmSSE{\sen}{\chi}{\varphi} \ldimp \mmSSE{\sen}{\lnot\chi}{\varphi}$ (communication on a topic is just as communication on its negation).

\ssparagraph{Axiom system.} The axioms and rule of \autoref{tbl:LOdSSE} form, together with those in \autoref{tbl:LOd}, a sound and strongly complete axiom system for \LAdSSE. They rely on the \ti{DEL} reduction axioms technique (for an explanation, see \cite{WangC13} or \cite[Section 7.4]{vanDitmarschEtAl2007}), with axiom \dsaSSE{\opgDist} being the crucial one. Using the abbreviation
\[
  \begin{array}{c}
    \mmDsr{\sag}{\chi}{\varphi} := (\chi \limp\mmDs{\sag}{(\chi \limp \varphi)}) \land (\lnot\chi \limp\mmDs{\sag}{(\lnot\chi \limp \varphi)}) \\
    \text{\small (\emph{``agents in $\sag$ know distributively that $\chi$'s truth value implies $\varphi$''}),}
  \end{array}
\]
the axiom indicates that a group $\sag$ knows $\varphi$ distributively after the action ($\mmSSE{\sen}{\chi}{\mmDg{\varphi}}$) if and only if the group $\sen \cup \sag$ knew, distributively, that $\varphi$ would hold after the action ($\mmDs{\sen \cup \sag}{\mmSSE{\sen}{\chi}{\varphi}}$) and the agents in $\sag$ know distributively that $\chi$'s truth-value implies the action will make $\varphi$ true ($\mmDsr{\sag}{\chi}{\mmSSE{\sen}{\chi}{\varphi}}$).

\begin{table}[t]
  \caption{Additional axioms and rules for \LOdSSE.}
  \renewcommand{\arraystretch}{1.5}
  \begin{smallctabular}{r@{\;\;\;}l}
    \toprule
    \dsaSSE{p}:        & $\vdash \mmSSE{\sen}{\chi}{p} \lldimp p$ \\
    \dsaSSE{\lnot}:    & $\vdash \mmSSE{\sen}{\chi}{\lnot \varphi} \lldimp \lnot \mmSSE{\sen}{\chi}{\varphi}$ \\
    \dsaSSE{\land}:    & $\vdash \mmSSE{\sen}{\chi}{(\varphi \land \psi)} \lldimp (\mmSSE{\sen}{\chi}{\varphi} \land \mmSSE{\sen}{\chi}{\psi})$ \\
    \dsaSSE{\opgDist}: & $\vdash \mmSSE{\sen}{\chi}{\mmDg{\varphi}} \lldimp (\mmDs{\sen \cup \sag}{\mmSSE{\sen}{\chi}{\varphi}} \land \mmDsr{\sag}{\chi}{\mmSSE{\sen}{\chi}{\varphi}})$ \\
    \dsreSSE:          & If $\vdash \varphi_1 \ldimp \varphi_2$ then $\vdash \mmSSE{\sen}{\chi}{\varphi_1} \ldimp \mmSSE{\sen}{\chi}{\varphi_2}$ \\
    \bottomrule
  \end{smallctabular}
  \label{tbl:LOdSSE}
\end{table}

From \autoref{tbl:LOdSSE} one can define a truth-preserving translation from \LAdSSE to \LAd, thanks to which the following theorem can be proved.

\begin{theorem}[\cite{Velazquez2022}]\label{teo:LOdSSE}
  The axiom system \LOdSSE (\LOd[\autoref{tbl:LOd}]+\autoref{tbl:LOdSSE}) is sound and strongly complete for \LAdSSE.
\end{theorem}

\nsparagraph{Structural equivalence.} The modality $\modboxSSE{\sen}{\chi}$ is invariant under collective bisimilarity.

\begin{theorem}[$\cobisim$ implies \LAdSSE-equivalence]\label{thm:cobisim.LAdSSE}
  Let $(M,w)$ and $(M', w')$ be two pointed models. If $(M,w) \cobisim (M',w')$ then, for every $\psi \in \LAdSSE$,
  \begin{ctabular}{c}
    $(M,w) \Vdash \psi$ \qtnq{if and only if} $(M',w') \Vdash \psi$.
  \end{ctabular}
  \begin{proof}
    The language \LAdSSE is \replace{the union of $\LAdSSE^i$ for all $i \in \Nat$}{$\bigcup_{i \in \Nat} \LAdSSE^i$}, so the proof proceeds by induction on $i$. \replace{In fact, the text will prove}{The text proves} a stronger statement: for every $\psi \in \LAdSSE$ and every $M$ and $M'$, if $(M,w) \cobisim (M',w')$ then \begin{inlineenumco} \item $(M,w) \Vdash \psi$ \replace{if and only if}{iff} $(M',w') \Vdash \psi$, and \item $(\mopSSE{M}{\sen}{\psi},w) \cobisim (\mopSSE{M'}{\sen}{\psi},w')$\end{inlineenumco}. Details can be found in the \hyperref[thm:cobisim.LAdSSE:proof]{appendix}.
  \end{proof}
\end{theorem}

\nsparagraph{Expressivity.}\label{par:trans.LAdSSE.to.LAd} It is clear that $\LAd \preccurlyeq \LAdSSE$, as every formula in the former is also in the latter. Moreover: the reduction axioms in \autoref{tbl:LOdSSE} define a recursive translation $\tr:\LAdSSE \to \LAd$ such that $\varphi \in \LAdSSE$ implies $\Vdash \varphi \ldimp \tr(\varphi)$ \cite{Velazquez2022}.\footnote{Note: the translation's complexity might be exponential, as it is for similar \ti{DEL}s (e.g., public announcement: \cite{Lutz2006}).} This implies $\LAdSSE \preccurlyeq \LAd$ and thus $\LAd \approx \LAdSSE$: the languages \LAd and \LAdSSE are equally expressive.

\msparagraph{Model checking} The original work on topic-based communication \cite{Velazquez2022} did not discuss computational complexity. Here we address that of the model checking problem for \LAdSSE.

Given a finite pointed model $(M,w)$ and a formula $\varphi \in \LAdSSE$, the model checking strategy is as follows. Start by creating the list $\sub(\varphi)$ of all subformulas of $\varphi$ and all partial communication modalities $\modboxSSE{\sen}{\chi}$ in it. Next, similarly to the approach in \cite{kuijer15}, label each element of $\sub(\varphi)$ with the sequence of partial communication modalities inside the scope of which it appears.  Finally, order the resulting \replace{labelled list}{list} \replace{in the following way}{as follows}: for $\psi^\sigma_1, \psi^\tau_2 \in \sub(\varphi)$ (with $\sigma$ and $\tau$ the labellings) we have that $\psi^\sigma_1$ precedes $\psi^\tau_2$ if and only if
\begin{compactitemize}
  \item $\psi^\sigma_1$ and  $\psi^\tau_2$ are parts of modalities $\modboxSSE{\sen}{\chi}$, and {$\sigma < \tau$},\footnote{That is, $\sigma$ is a proper prefix of $\tau$.} or else
  \item $\psi^\sigma_1$ appears within some $\modboxSSE{\sen}{\chi}$, and $\psi^\tau_2$ does not, or else
  \item  $\psi^\sigma_1$ is of the form $\modboxSSE{\sen}{\chi}$, $\psi^\tau_2$ is not,  and $\sigma < \tau$, or else
  \item neither $\psi^\sigma_1$ nor $\psi^\tau_2$ are parts of some $\modboxSSE{\sen}{\chi}$, and $\tau < \sigma$, or else
  \item both $\psi^\sigma_1$ are $\psi^\tau_2$ are of the form $\modboxSSE{\sen}{\chi}$, and $\sigma < \tau$, or else
  \item $\sigma = \tau$, and $\psi^\sigma_1$ is a part of $\psi^\tau_2$, or else
  \item $\psi_1$ appears to the left of $\chi$ in $\varphi$.
\end{compactitemize}
The intuition behind such an ordering is to allow a model checking algorithm to deal with $\chi$'s within $\modboxSSE{\sen}{\chi}$'s before dealing with formulas within the scope of the modality. This way we ensure that, when we need to evaluate $\varphi$ in $\modboxSSE{\sen}{\chi} \varphi$, we already know the effect of $\modboxSSE{\sen}{\chi}$ on the model. As an example, consider $\varphi := \modboxSSE{\sen_1}{p \land q}\modboxSSE{\sen_2}{q} \mmDs{\sag}{p}$. The resulting ordered list $\sub(\varphi)$ is then $p$, $q$, $p \land q$, $\modboxSSE{\sen_1}{p \land q}$, $q^{\modboxSSE{\sen_1}{p \land q}}$, $\modboxSSE{\sen_2}{q}^{\modboxSSE{\sen_1}{p \land q}}$, $p^{\modboxSSE{\sen_1}{p \land q}, \modboxSSE{\sen_2}{q}}$, $\mmDs{\sag}{p}^{\modboxSSE{\sen_1}{p \land q}, \modboxSSE{\sen_2}{q}}$, $\modboxSSE{\sen_2}{q} \mmDs{\sag}{p}^{\modboxSSE{\sen_1}{p \land q}}$, $\varphi$.

\replace{Observe that}{Note:} each subformula of $\varphi$ is labelled with exactly one (maybe empty) sequence of partial communication modalities.  Moreover, we label communication modality symbols separately. The number of subformulas of $\varphi$ and modality symbols is bounded by $\mathcal{O}(|\varphi|)$.  Since each element of $\sub(\varphi)$ is labelled by only one sequence of modalities, we use at most polynomial number of them.

\begin{center}
  \begin{breakablealgorithm}
    \caption{An algorithm for global model checking for \LAdSSE{}}\label{euclid2}
    \footnotesize
    \begin{algorithmic}[1]
      \Procedure{GlobalMC}{$M, \varphi$}
      \ForAll{$\psi^\sigma \in \sub(\varphi)$}
        \ForAll{$w \in W$}
      \Case{$\psi^\sigma = \mmDs{\sag}{\chi}^\sigma$}       
        \State{$\mathit{check} \gets \mathit{true}$}
        \ForAll{$(w,v) \in \R_\sag$}
          \If{$(w,v)$ is labelled with $\sigma$}
          
          \If{$v$ is not labelled with $\chi^\sigma$}
            \State{$\mathit{check} \gets \mathit{false}$}
            \State{\textbf{break}}
          \EndIf
          \EndIf
        \EndFor
        \If{$\mathit{check}$}
          \State{label $w$ with $\mmDs{\sag}{\chi}^\sigma$}
        \EndIf
      \EndCase
        
          \Case{$\psi^\sigma = \modboxSSE{\sen}{\chi}^\sigma$}
            \ForAll{$\agi \in \ag$}
              \ForAll{$(v,u) \in \R_\agi$}
                \If{$(v,u)$ is labelled with $\sigma$}
                  \If{$v$ is labelled with $\chi$ iff $u$ is labelled with $\chi$}
                    \State{label $(v,u)$ with $\sigma, \modboxSSE{\sen}{\chi}$}
                    \Else
                    \State{$\mathit{check} \gets \mathit{true}$}
                    \ForAll{$\agj \in \sen$}
                      \If{$(v,u) \not \in \R_\agj$}
                        \State{$\mathit{check} \gets \mathit{false}$}
                        \State{\textbf{break}}
                      \EndIf
                    \EndFor
                    \If{$\mathit{check}$}
                      \State{label $(v,u)$ with $\sigma, \modboxSSE{\sen}{\chi}$}
                    \EndIf
                  \EndIf
                \EndIf
              \EndFor
            \EndFor
          \EndCase
          
          \Case{$\psi^\sigma = \modboxSSE{\sen}{\chi} \xi^\sigma$}
            \If{$w$ is labelled with $\xi^{\sigma, \modboxSSE{\sen}{\chi}}$}
              \State{label $w$ with $\modboxSSE{\sen}{\chi} \xi^\sigma$}
            \EndIf
          \EndCase
        \EndFor
    \EndFor      
      
     \EndProcedure
    \end{algorithmic}
  \end{breakablealgorithm}
\end{center}

The labelling Algorithm \ref{euclid2} is inspired by \replace{the algorithm}{that} for epistemic logic \cite{halpern92}. The crucial difference is that, besides labelling states, we also label transitions (case $\modboxSSE{\sen}{\chi}^\sigma$). This allows us to keep track of which relations `survive' updates with partial communication modalities. The labelling of transitions is taken into account when dealing with the epistemic case $\mmDs{\sag}{\chi}^\sigma$: we check only transitions that have `survived' at a current step of an algorithm run. 

Correctness of the algorithm can be shown by an induction on $\varphi$, noting that cases of the algorithm mimic the definition of semantics. From a computational perspective, the preparation of $\sub(\varphi)$ can be done in $\mathcal{O}(|\varphi|^2)$ steps. The running time of \textsc{GlobalMC} is bounded by $\mathcal{O}(|\varphi|^2 \cdot |W| \cdot |\ag| \cdot |R|)$ for the case of $\modboxSSE{\sen}{\chi}^\sigma$. 

\begin{theorem}\label{thm:LAdSSE.mc}
  The model checking problem for \LAdSSE is in \textit{P}.
\end{theorem}

\section{Partial communication vs. public announcements}\label{sec:SSE.vs.PAL}

The action for partial communication is, in a sense, similar to that for a public announcement: both are epistemic actions through which agents receive information about the truth-value of a specific formula. The difference is that, while in the latter the information comes from an external source, in the former the information comes from agents in the model. It makes sense to discuss the relationship between their formal representations.

Under its standard definition \cite{Plaza1989}, the public announcement of a formula $\xi$ transforms a model by eliminating all $\lnot\xi$-worlds. For a fair comparison with partial communication, here is an alternative public announcement definition that rather removes all edges between worlds disagreeing on $\xi$'s truth-value \cite{BenthemL07}.\footnote{Cf. \cite{GerbrandyGroeneveld1997}, which removes only edges pointing to $\lnot\xi$-worlds. The option used here has the advantage of behaving, with respect to the preservation of certain relational properties (reflexivity, symmetry, transitivity), as the standard definition does.}

\begin{definition}[Public announcement]\label{def:PAmo}
  Let $M = \tupla{W, \R, V}$ be a model; take a formula $\xi$. The model $\mopPA{M}{\xi} = \tuplan{W, \relPA{\R}{\xi}, V}$ 
  is such that
  \begin{ctabular}{c}
    $\sbi{\relPA{\R}{\xi}} := \sbi{\R} \cap {\knonfu{M}{\xi}}$.
  \end{ctabular}
  Thus, $\sbDg{\relPA{R}{\xi}} = \sbDs{R}{\sag} \cap {\knonfu{M}{\xi}}$.
\end{definition}

The world-removing version and the edge-deleting alternative are collectively $\pa$-bisimilar (see \nameref{pro:wdPA.eq.RDpa} in the appendix), and thus interchangeable from \LAd's perspective. Here is a modality for describing the operation's effect.

\begin{definition}[Modality $\modboxPA{\xi}$]\label{def:PAmm}
  The language \LAdPA extends \LAd with a modality $\modboxPA{\xi}$ for $\xi$ a formula.\footnote{\label{ftn:def:PAmm}More precisely, $\LAdPA^1$ extends $\LAdPA^0 = \LAd$ with \replace{an additional modality}{} $\modboxPA{\xi}$ for $\xi \in \LAdPA^0$, $\LAdPA^2$ extends $\LAdPA^1$ with \replace{an additional modality}{} $\modboxPA{\xi}$ for $\chi \in \LAdSSE^1$ and so on. The language \LAdPA is \replace{the union of all $\LAdPA^i$ with $i \in \Nat$}{then defined as $\bigcup_{i \in \Nat} \LAdPA^i$}.} For their semantic interpretation,
  \begin{lbtabular}{l@{\qssidefq}l}
    $(M, w) \Vdash \mmPA{\xi}{\varphi}$ & $(M, w) \Vdash \xi$ \;implies\; $(\mopPA{M}{\xi}, w) \Vdash \varphi$.
  \end{lbtabular}
  Defining $\mmdiaPA{\xi}{\varphi} := \lnot \mmPA{\xi}{\lnot \varphi}$ implies $\Vdash \mmdiaPA{\xi}{\varphi} \ldimp (\xi \land \mmPA{\xi}{\varphi})$.
\end{definition}

It can be shown that \LAdPA is invariant under collective bisimilarity (see \nameref{thm:cobisim.LAdPA} in the appendix). An axiom system can be obtained by using the reduction axioms technique, with the crucial axiom being $\mmPA{\xi}{\mmDg{\varphi}} \lldimp (\xi \limp \mmDg{\mmPA{\xi}{\varphi}})$ \cite{WangA13}. As before, the existence of the reduction axioms implies $\LAdPA \preccurlyeq \LAd$. This, together with the straightforward $\LAd \preccurlyeq \LAdPA$, implies $\LAd \approx \LAdPA$: the languages \LAd and \LAdPA are equally expressive.

When comparing partial communication with public announcements, a first natural question is about the languages' relative expressivity. The answer is simple: \LAdSSE and \LAdPA are both reducible to \LAd, and thus they are equally expressive.

At the semantic level, one might wonder whether the operations can `mimic' each other. More precisely, one can ask the following.
\begin{compactitemize}
  \item Given $\xi \in \LAd$: are there $\sen \subseteq \ag$, $\chi \in \LAd$ such that $\mopPA{M}{\xi} \cobisim \mopSSE{M}{\sen}{\chi}$ for every $M$?
   (In symbols: $\forall \xi \,.\, \exists \sen \,.\, \exists \chi \,.\, \forall M \,.\, (\mopPA{M}{\xi} \cobisim \mopSSE{M}{\sen}{\chi})$?)
  \item Given $\sen \subseteq \ag$, $\chi \in \LAd$: is there $\xi \in \LAd$ such that $\mopSSE{M}{\sen}{\chi} \cobisim \mopPA{M}{\xi}$ for every $M$?
   (In symbols: $\forall \sen \,.\, \forall \chi \,.\, \exists \xi \,.\, \forall M \,.\, (\mopSSE{M}{\sen}{\chi} \cobisim \mopPA{M}{\xi})$?)
\end{compactitemize}
Some known model-update operations have this relationship. For example, action models \cite{BaltagMS98} generalise a standard public announcement: for every formula $\xi$ there is an action model that, when applied to any model, produces exactly the one that a public announcement of $\xi$ does. For another example, edge-deleting versions of a public announcement (both that in \cite{GerbrandyGroeneveld1997} and that in Definition \ref{def:PAmo}) can be represented within the arrow update framework of \cite{KooiR11}.

Here, the answer to the first question is straightforward: the agents might not have, even together, the information that a public announcement provides.

\begin{fact}\label{fct:no.pc}
  Take $\ag = \set{\aga}$ and $\pa = \set{p}$; consider the (reflexive and symmetric) model $M$ below on the left. A public announcement of $p$ yields the model on the right.
  \begin{ctabular}{c@{\qquad}c@{\qquad}c}
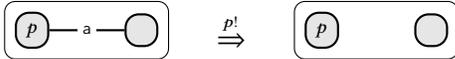
    \begin{tabular}{@{}c@{}}
      \begin{tikzpicture}[frame rectangle]
        \node [myworld] (w0) {$p$};
        \node [myworld, right = of w0] (w1) {};

        \path (w0) edge [myarrow, -] node [myarrow-label, within] {$\aga$} (w1);
      \end{tikzpicture}
    \end{tabular}
    &
    \begin{tabular}{@{}c@{}}
      {\Large $\overset{\mopPA{}{p}}{\Rightarrow}$}
    \end{tabular}
    &
    \begin{tabular}{@{}c@{}}
      \begin{tikzpicture}[frame rectangle]
        \node [myworld] (w0) {$p$};
        \node [myworld, right = of w0] (w1) {};
      \end{tikzpicture}
      \end{tabular}
  \end{ctabular}
  Now, there is no $\sen \subseteq \ag$ and $\chi \in \LAd$ such that $\mopSSE{M}{\sen}{\chi} \cobisim \mopPA{M}{p}$. The group $\sen$ can be only $\emptyset$ or $\set{\aga}$ and, in both cases, $\sba{\relSSE{\R}{\sen}{\chi}} = \sba{\R}$, regardless of the formula $\chi$.
\end{fact}

Thus, $\forall M \,.\, \forall \xi \,.\, \exists \sen \,.\, \exists \chi \,.\, (\mopPA{M}{\xi} \cobisim \mopSSE{M}{\sen}{\chi})$ fails: for the given model, the effect of a public announcement of $p$ cannot be replicated by any act of partial communication. This answers negatively the (stronger) first question above: there are no agents $\sen$ and topic $\chi$ that can replicate the given public announcement in every model.

The answer to the second question is interesting: through partial communication, the agents can reach epistemic states that cannot be reached by a public announcement.

\begin{fact}\label{fct:no.pa}
  Take $\ag = \set{\aga, \agb}$ and $\pa = \set{p,q}$; consider the (reflexive and symmetric) model $M$ below on the left. A partial communication between all agents about $p \ldimp q$ (equivalence classes highlighted) yields the model on the right.
  \begin{ctabular}{@{}c@{\quad}c@{\quad}c@{}}
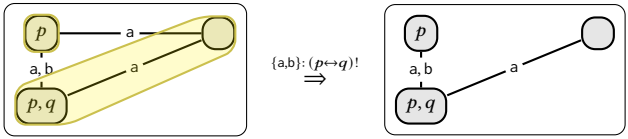
    \begin{tabular}{@{}c@{}}
      \begin{tikzpicture}[frame rectangle, node distance = 1.5em and 6em]
        \node [myworld] (w0) {$p$};
        \node [myworld, right = of w0] (u1) {};
        \node [myworld, below = of w0] (u0) {$p,q$};

        \path (w0) edge [myarrow, -] node [myarrow-label, within] {$\aga, \agb$} (u0)
                   edge [myarrow, -] node [myarrow-label, within] {$\aga$} (u1)
              (u0) edge [myarrow, -] node [myarrow-label, within] {$\aga$} (u1);

        \draw[topicclass] (w0.north west) -- (w0.north east) -- (w0.south east) -- (w0.south west) -- cycle;
        \draw[topicclass] (u1.north west) -- (u1.north east) -- (u1.south east) -- (u0.south east) -- (u0.south west) -- (u0.north west) -- cycle;
      \end{tikzpicture}
    \end{tabular}
    &
    \begin{tabular}{@{}c@{}}
      $\overset{\mopSSE{}{\set{\aga, \agb}}{(p \ldimp q)}}{\Rightarrow}$
    \end{tabular}
    &
    \begin{tabular}{@{}c@{}}
      \begin{tikzpicture}[frame rectangle, node distance = 1.5em and 6em]
        \node [myworld] (w0) {$p$};
        \node [myworld, right = of w0] (u1) {};
        \node [myworld, below = of w0] (u0) {$p,q$};

        \path (w0) edge [myarrow, -] node [myarrow-label, within] {$\aga, \agb$} (u0)
              (u0) edge [myarrow, -] node [myarrow-label, within] {$\aga$} (u1);
      \end{tikzpicture}
    \end{tabular}
  \end{ctabular}
  Now, there is no $\xi \in \LAd$ such that $\mopPA{M}{\xi} \cobisim \mopSSE{M}{\set{\aga, \agb}}{(p \ldimp q)}$. For this, note that a public announcement preserves transitive indistinguishability relations; yet, while $M$ is transitive, $\mopSSE{M}{\set{\aga, \agb}}{(p \ldimp q)}$ is not.
\end{fact}

Thus, $\forall M \,.\, \forall \sen \,.\, \forall \chi \,.\, \exists \xi \,.\, (\mopSSE{M}{\sen}{\chi} \cobisim \mopPA{M}{\xi})$ fails: for the provided model, the effect of a `conversation' among $\aga$ and $\agb$ on $p \ldimp q$ cannot be replicated by any public announcement. This answers negatively the (stronger) second question above: there is no $\chi$ that can replicate the given partial communication in every model.

\section{Arbitrary partial communication}\label{sec:SAE}

The partial communication framework allows us to model inter-agent information exchange. Yet, consider competitive scenarios. While it is interesting to find out what a form of partial communication can achieve (fix the agents and the topic, then find the consequences), one might be also interested in deciding whether a given goal can be achieved by \emph{some} form of partial communication (fix the \emph{goal}: is there a group of agents and a topic that can achieve it?). This \emph{quantification} over the sharing agents and the topic they discuss adds a \emph{strategic} dimension to the framework. This is particularly useful when communication occurs over an insecure channel, as one would like to know \emph{whether} some form of partial communication (who talks, and on which topic) can achieve a given goal (e.g., make something group or common knowledge while also precluding adversaries or eavesdroppers from learning it, as in \cite{Ditmarsch03}). Thus, in the spirit of \cite{BalbianiBDHHL08}, one can then \emph{quantify}, either over the agents that communicate or over the topic they discuss.

\smallskip

Quantifying over the communicating agents does not need additional machinery: $\ag$ is finite, so a modality stating that \emph{``$\varphi$ is true after any group of agents share all their information about $\chi$''} is definable as $\mmSSE{\ast}{\chi}{\varphi} := \bigwedge_{\sen \subseteq \ag} \mmSSE{\sen}{\chi}{\varphi}$. Quantifying over the topic, though, requires additional tools.

\subsection{Language, semantics, and basic results}

\begin{definition}[Modality $\modboxSAE{\sen}$]\label{def:SAEmm}
  The language \LAdSAE{} extends \LAdSSE with a modality $\modboxSAE{\sen}$ for each group of agents $\sen \subseteq \ag$. More precisely, take $\LAdSAE{,0} = \LAd^\ast$ to be \LAd plus \replace{the modality}{} $\modboxSAE{\sen}$. Then, define $\LAdSAE{,i+1}$ as the result of extending $\LAdSAE{,i}$ with \replace{an additional modality}{} $\modboxSSE{\sen}{\chi}$ for $\sen \subseteq \ag$ and $\chi \in \LAdSAE{,i}$. The language \LAdSAE{} is \replace{the union of all $\LAdSAE{,i}$ with $i \in \Nat$}{defined as $\bigcup_{i \in \Nat} \LAdSAE{,i}$}.   
  For the semantic interpretation,
  \begin{lbtabular}{@{}l@{\;\;\ssidef\;\;}l@{}}
    $(M, w) \Vdash \mmSAE{\sen}{\varphi}$ & every $\chi \in \LAd$ is s.t. $(\mopSSE{M}{\sen}{\chi}, w) \Vdash \varphi$ \\
    \multicolumn{1}{l}{}                  & (every $\chi \in \LAd$ is s.t. $(M, w) \Vdash \mmSSE{\sen}{\chi}{\varphi}$).
  \end{lbtabular}
  If one defines $\mmdiaSAE{\sen}{\varphi} := \lnot \mmSAE{\sen}{\lnot \varphi}$, then
  \begin{lbtabular}{@{}l@{\;\;\ssidef\;\;}l@{}}
    $(M, w) \Vdash \mmdiaSAE{\sen}{\varphi}$ & there is $\chi \in \LAd$ s.t. $(\mopSSE{M}{\sen}{\chi}, w) \Vdash \varphi$.
  \end{lbtabular}
  The size of $\varphi \in \LAdSAE{}$ is defined as in \autoref{def:SSEmm} with the following additional clause: $\card{\mmSAE{\sen}{\varphi}} := \card{\varphi} + 1$. 
\end{definition}

Note: $\modboxSAE{\sen}$ quantifies over formulas in \LAd, and not over formulas in \LAdSAE{}. As in \cite{BalbianiBDHHL08}, this is to avoid circularity issues. One could have also chosen to quantify over formulas in \LAdSSE, but $\LAd \approx \LAdSSE$ (\autopageref{par:trans.LAdSSE.to.LAd}) so nothing is lost by using \LAd instead.\footnote{Still, for languages with other types of group knowledge, adding a dynamic modality might increase the expressive power. For more on this (in the context of common knowledge and quantified announcements), the reader is referred to \cite{GalimullinA21}.}

\msparagraph{Axiom system.} Axiomatising \LAdSAE{} requires an additional notion.

\begin{definition}[Necessity Forms]
  Take $\varphi \in \LAdSAE{}$, $\chi \in \LAd$, $\sen, \sag \subseteq \ag$ and $\sharp \not \in P$. The set of \emph{necessity forms} \cite{goldblatt82} is given by
  \[
    \eta(\sharp)::= \sharp \mid \varphi \rightarrow \eta(\sharp) \mid \mmDg{\eta(\sharp)} \mid \mmSSE{\sen}{\chi}{\eta(\sharp)}
  \]
  The result of replacing $\sharp$ with $\varphi$ in \replace{a necessity form}{} $\eta(\sharp)$ is denoted as $\eta(\varphi)$.
\end{definition}

The (note: \emph{infinitary}) axiom system for \LAdSAE{} is given by the axioms and rules on Tables \ref{tbl:LOd}, \ref{tbl:LOdSSE} and \ref{tbl:LOdSAE}. The system is similar to well-known axiomatisations of other logics of quantified epistemic actions (see \cite{vanditmarsch20} for an overview). In \autoref{tbl:LOdSAE}, the soundness of \ax{A}{}{\modboxSAE{\sen}} and \ax{R}{}{\modboxSAE{\sen}} follow from $\modboxSAE{\sen}$'s semantic interpretation. Completeness of the whole system can be shown by combining and adapting techniques from \cite{WangA13} (to deal with distributed knowledge) and \cite{balbiani15} (to tackle quantifiers). The reader interested in details is referred to \cite{agotnes22}, where the authors presented a relatively similar completeness proof for a system with distributed knowledge and quantification over public announcements.

\begin{table}[t]
  \renewcommand{\arraystretch}{1.5}
  \caption{Axiom and rule of inference for the arbitrary case.}
  \begin{smallctabular}{r@{\;\;\;}l}
    \toprule
    \ax{A}{}{\symbSAE{\sen}}: & $\vdash \mmSAE{\sen}{\varphi} \rightarrow \mmSSE{\sen}{\chi}{\varphi}$ \quad for $\chi \in \LAd$ \\
    \ax{R}{}{\symbSAE{\sen}}: & If $\vdash \eta(\mmSSE{\sen}{\chi}{\varphi})$ for all $\chi \in \LAd$, then $\vdash \eta(\mmSAE{\sen}{\varphi})$ \\
    \bottomrule
  \end{smallctabular}
  \label{tbl:LOdSAE}
\end{table}

\begin{theorem}\label{teo:LOdSAE}
  The axioms and rules on Tables \ref{tbl:LOd}, \ref{tbl:LOdSSE} and \ref{tbl:LOdSAE} are sound and (together) complete for \LAdSAE{}.
\end{theorem}

\nsparagraph{Structural equivalence.} The modality $\modboxSAE{\sen}{}$ is also invariant under collective bisimilarity.

\begin{theorem}[$\cobisim$ implies \LAdSAE{}-equivalence]\label{thm:cobisim.LAdSAE}
  Let $(M,w)$ and $(M', w')$ be two pointed models. If $(M,w) \cobisim (M',w')$ then, for every $\psi \in \LAdSAE{}$,
  \begin{ctabular}{c}
    $(M,w) \Vdash \psi$ \qtnq{if and only if} $(M',w') \Vdash \psi$.
  \end{ctabular}
  \begin{proof}
    As for \autoref{thm:cobisim.LAdSSE} (see the \hyperref[thm:cobisim.LAdSAE:proof]{appendix}).
  \end{proof}
\end{theorem}

\nsparagraph{Expressivity.}\label{par:LAdSSE.vs.LAdSAE} The modality $\modboxSAE{\sen}$ adds expressive power.

\begin{theorem}
  \LAdSAE{} is strictly more expressive than \LAdSSE.
\end{theorem}

This result can be proven as the analogous result for APAL \cite[Proposition 3.13]{BalbianiBDHHL08}. Assume towards a contradiction that the languages are equally expressive so, given a formula in \LAdSAE{}, there is an equivalent formula in \LAdSSE. Since both formulas are finite, there is an atom $p$ that appears in neither. However, $\modboxSAE{\sen}$ in \LAdSAE{} quantifies over any formula, and thus over formulas including $p$. With this, one can build two models that include worlds that satisfy $p$. Then, using induction, we can show that the formula in \LAdSSE (without $p$) cannot tell the models apart, while the formula in \LAdSAE{} (where quantification ranges also over formulas with $p$) can. This technique is used (with more details) in the proofs in Section \ref{sbs:apc.vs.apal}.


\subsection{Model checking}

\replace{Here it is shown that the}{The} complexity of the model checking problem for \LAdSAE{} is \textit{PSPACE}-complete\replace{. This result}{: this} is in line with \replace{the \textit{PSPACE}-completeness of many}{the complexity of} other logics of quantified information change as, e.g., arbitrary public announcements \cite{BalbianiBDHHL08}, group announcement logic \cite{agotnes10}, coalition announcement logic \cite{alechina21} and arbitrary arrow update logic \cite{vanditmarsch17}. However, \replace{there is an interesting twist in our algorithm}{this case has an interesting twist}. Model checking algorithms for the aforementioned logics \replace{include a step of computing}{compute} a bisimulation contraction of the model, and then continue working on the contracted model. This is not possible in our case: a model and its \emph{collective} bisimulation contraction are not collectively bisimilar \cite{Roelofsen2005}: they might differ in some formulas' truth-value. We still compute bisimulation contractions, but we use them just to inform our algorithm about bisimilar states. The computation continues on the original non-contracted model. 
 
\begin{definition}[$\sen$-definable restrictions]
  Let $(M,w)$ be a pointed model; take $\sen \subseteq \ag$. A model $(N,w)$ is an $\sen$-definable restriction of $(M,w)$ if and only if $(N,w) = (\mopSSE{M}{\sen}{\chi},w)$ for some $\chi \in \LAdSAE{}$. 
\end{definition}

\begin{fact}
Let $(M,w)$ be a finite pointed model. Then there is a finite number of $\sen$-definable restrictions of $(M,w)$.
\end{fact}

The proof below presents an algorithm $\mc(M,w,\varphi)$ that returns \emph{true} if and only if $(M,w) \Vdash \varphi$, and returns \emph{false} if and only if $(M,w) \not \Vdash \varphi$. The main challenge is that modalities $\mmSAE{\sen}$ quantify over an \emph{infinite} number of formulas. However, for any given \emph{finite} model $M$, there is only a \emph{finite} number of possible $\sen$-definable model restrictions. Showing that the problem is \ti{PSPACE}-hard uses the classic reduction from the satisfiability of QBF.


\begin{theorem}\label{thm:LAdSAE.mc}
  The model checking for \LAdSAE{} is \ti{PSPACE}-complete.
  \begin{proof}
    Let $(M,w)$ be a pointed model, and $\varphi \in  \LAdSAE{}$. In \autoref{euclid}, Boolean cases and the case for $\modDg$ are as expected, and thus omitted.
    \begin{center}
      \begin{breakablealgorithm}
        \caption{An algorithm for model checking for \LAdSAE{}}\label{euclid}
        \footnotesize
        \begin{algorithmic}[1]
          \Procedure{MC}{$M, w, \varphi$}
          \Case {$\varphi =  \mmSSE{\sen}{\chi}{\psi}$}
          \State{\tb{return} $\textsc{MC}(\mopSSE{M}{\sen}{\chi}, w, \psi)$}
          \EndCase

          \Case {$\varphi = \mmSAE{\sen}{\psi}$}
          \State{Compute collective $\pa$-bisimulation contraction $\|M\|^C$}
          \ForAll{$\sen$-definable restrictions $(N, w)$ of $(M, w)$}
            \If{$\textsc{MC}(N, w, \psi)$ \tb{returns} \ti{false}}
            \State{\tb{return} \ti{false}}
            \EndIf
          \EndFor
          \State{\tb{return} \ti{true}}
          \EndCase
         \EndProcedure
        \end{algorithmic}
      \end{breakablealgorithm}
    \end{center}
    The basic idea in the construction of $\sen$-definable restrictions is to consider a subset of all possible bipartitions of $(M,w)$, taking care that bisimilar states end up in the same partition. This can be done by checking that for each state, if it is in a partition, then all states in the same collective bisimulation equivalence class are also in the same partition. Collective bisimulation equivalence classes can be computed by, e.g., a modification of Kanellakis-Smolka algorithm \cite{kanellakis90} that takes into account not only relations but also intersections thereof. Having computed collective bisimulation equivalence classes of $(M,w)$, one can construct an $\sen$-definable restriction of the model by taking a bipartition such that if $v$ belongs to one partition, then all $u \in [v]$ also belong to the same partition, with $[v]$ being a collective bisimulation equivalence class.

    Constructing restrictions takes polynomial time and thus space. The space required for the case of $\mmSSE{\sen}{\chi}{\psi}$ is bounded by $\mathcal{O}(|\varphi| \cdot |M|)$. For the case of $\mmSAE{\sen}{\psi}$, collective bisimulation contraction can be computed in polynomial time and space, and each restriction has a size of at most $|M|$. If one traverses a given formula depth-first and reuses memory, the space to store model restrictions is polynomial in $|\varphi|$ (even though the algorithm itself runs in exponential time). Thus, the space required for the case of $\mmSAE{\sen}{\psi}$ is bounded by $\mathcal{O}(|\varphi| \cdot |M|)$.

    Finally, since computing each subformula of $\varphi$ requires space bounded by $\mathcal{O}(|\varphi| \cdot |M|)$, the space required by the whole algorithm is bounded by $\mathcal{O}(|\varphi|^2 \cdot |M|)$. The algorithm follows closely the semantics of \LAdSAE{}, and correctness can be shown via induction on $\varphi$. For the case of quantifiers note that, in order to switch from bipartitions to particular formulas corresponding to those partitions, one can use characteristic formulas \cite{vanditmarsch14}. These formulas are built in such a way that they are true only in one state of a model (up to collective bisimularity). 

    \smallskip

    \replace{To show that the model checking problem is \ti{PSPACE}-hard}{For showing \ti{PSPACE}-hardness}, use the classic reduction from the satisfiability of QBF. W.l.o.g., consider QBFs without free variables in which every variable is quantified only once. Consider a QBF with $n$ variables $\set{x_1, \ldots, x_n}$. We need a model and a formula in \LAdSAE{} that are both of polynomial size of the QBF.  The (reflexive and symmetric) model $M^n$ below satisfies this: $w_0$ is the evaluation point, and for each variable $x_i$ there are two states, $w_i^1$ and $w_i^0$, corresponding respectively to evaluating $x_i$ to $1$ and to $0$. Assume that each $w_i^1$ satisfies only $p_i$ and each $w_i^0$ satisfies only $q_i$. 
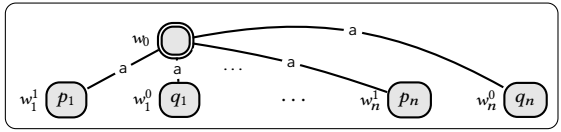
    \begin{center}
      \begin{tikzpicture}[node distance = 1em and 3em, frame rectangle]
        \node [myworld, label = {[myworld-label, outer sep = 2pt]left:$w_0$}, double] (w0) {};
        \node [myworld, label = {[myworld-label]left:$w_1^1$}, below left = of w0] (w11) {$p_1$};
        \node [myworld, label = {[myworld-label]left:$w_1^0$}, right = of w11] (w10) {$q_1$};
        \node [right = of w10] (w) {$\ldots$};
        \node [myworld, label = {[myworld-label]left:$w_n^1$}, right = of w] (wn1) {$p_n$};
        \node [myworld, label = {[myworld-label]left:$w_n^0$}, right = of wn1] (wn0) {$q_n$};

        \path (w0) 
                   edge [myarrow, -] node [myarrow-label, within] {$\aga$} (w11)
                   edge [myarrow, -] node [myarrow-label, within] {$\aga$} (w10)
                   edge [myarrow, draw = white] node [myarrow-label, within] {$\ldots$} (w)
                   edge [myarrow, -, bend left = 5pt] node [myarrow-label, within] {$\aga$} (wn1)
                   edge [myarrow, -, bend left = 20pt] node [myarrow-label, within] {$\aga$} (wn0);
      \end{tikzpicture}
    \end{center}
    Let $\Psi := Q_1 x_1 \mathellipsis Q_n x_n \Phi(x_1, \mathellipsis, x_n)$ be a quantified Boolean formula (so $Q_i \in \{\forall, \exists\}$ and $\Phi(x_1, \mathellipsis, x_n)$ is Boolean). The formula $\mathit{chosen}_k$ below indicates, intuitively, that the values (either 1 or 0) of the first $k$ variables have been chosen.
    \[
      \mathit{chosen}_k
      :=
      \bigwedge_{1 \leqslant i \leqslant k} (\mmKDa{p_i} \leftrightarrow \lnot \mmKDa{q_i})
      \land
      \bigwedge_{k < i \leqslant n} (\mmKDa{p_i} \land \mmKDa{q_i}).
    \]
    Here is, then, a recursive translation from a QBF $\Psi$ to a formula $\psi$ in \LAdSAE{}: $\psi_0 := \Phi(\mmKDa{p_1}, \mathellipsis,  \mmKDa{p_n})$,
    \begin{ctabular}{@{}r@{\,:=\,}l@{\;\quad\;}r@{\,:=\,}l@{\;\quad\;}r@{\,:=\,}l@{}}
      $\psi_k$ & $\left\{ \begin{array}{ll}
                            \modboxSAE{\set{\aga, \agb}} (\mathit{chosen}_k \rightarrow \psi_{k-1}) & \text{if } Q_k = \forall \\
                            \moddiaSAE{\set{\aga, \agb}} (\mathit{chosen}_k \land \psi_{k-1})       & \text{if } Q_k = \exists
                          \end{array} \right.$,
    \end{ctabular}
    $\psi := \psi_{n}$.
    We need to show that
    \begin{center}
      $Q_1 x_1 \mathellipsis Q_n x_n \Phi(x_1, \mathellipsis, x_n)$ is satisfiable \quad{if and only if}\quad $(M^n,w_0) \Vdash \psi$.
    \end{center}
    For this, observe that each state in $M^n$ can be characterised by a unique formula. Moreover, relation $\agb$ is the identity. Therefore, $\modboxSAE{\set{\aga, \agb}}$ and $\mmdiaSAE{\set{\aga, \agb}}$ can force any restriction of $\aga$-arrows from $w_0$ to $w_i$'s.  In the model, states $w^1_i$ and $w^0_i$ correspond the truth-value of $x_i$. The guard $\mathit{chosen}_k$ guarantees that only the truth-values of the first $k$ variables have been chosen, and that they have been chosen unambiguously (i.e. there is exactly one edge from $w_0$ to either $w^1_i$ and $w^0_i$). Thus, together with $\modboxSAE{\set{\aga, \agb}}$ and $\mmdiaSAE{\set{\aga, \agb}}$, the guards $\mathit{chosen}_k$ emulate $\forall$ and $\exists$. Then, once the values of all $x_i$'s have been set,  the evaluation of the QBF corresponds to the $\aga$-reachability of the corresponding states in $M^n$.
  \end{proof}
\end{theorem}

\subsection{Arbitrary partial communication vs. arbitrary public announcements}\label{sbs:apc.vs.apal}

The languages \LAdSSE and \LAdPA are equally expressive (both `reduce' to \LAd). As it is shown below, this changes when quantification (over topics and announced formulas, respectively) is added.

\begin{definition}
  The language \LAdAPA extends \LAdPA with a modality $\modboxPA{\ast}$ such that
  \begin{lbtabular}{l@{\;\;\ssidef\;\;}l@{}}
    $(M, w) \Vdash \mmPA{\ast}{\varphi}$ & for every $\chi \in \LAd$: $(M, w) \Vdash \mmPA{\chi}{\varphi}$.\footnotemark
  \end{lbtabular}
  \footnotetext{Thus, \LAdAPA extends the language from \cite{BalbianiBDHHL08} with the distributed knowledge modality.}
  Define $\mmdiaPA{\ast}{\varphi} := \lnot \mmPA{\ast}{\lnot \varphi}$, as usual.
\end{definition}

The theorem below shows that \LAdAPA and \LAdSAE{} are incomparable w.r.t. expressive power (i.e., $\LAdSAE{} \not\preccurlyeq \LAdAPA$ and $\LAdAPA \not\preccurlyeq \LAdSAE{}$). This result is obtained by adapting techniques and models from \cite{BalbianiBDHHL08} and \cite{vanditmarsch17} to the case of partial communication.\footnote{For space reasons, we do not present the whole argument here.}

\begin{theorem}
  \LAdAPA and \LAdSAE{} are, expressivity-wise, incomparable.
  \begin{proof}
    For $\LAdSAE{} \not\preccurlyeq \LAdAPA$, consider $\mmdiaSAE{\set{\aga, \agb}}{(\mmKb{p} \land \lnot \mmKb{\mmKb{p}})}$ in $\LAdSAE{}$. For a contradiction, assume there is an equivalent $\alpha \in \LAdAPA$. Since $\alpha$ is finite, there is an atom $q$ that does not occur in it. The strategy consists in building two $\pa \setminus \setn{q}$-bisimilar pointed models, then argue that they can be distinguished by $\mmdiaSAE{\set{\aga, \agb}}{(\mmKa{p} \land \lnot \mmKa{\mmKa{p}})}$ but not by any $\alpha$. Consider the (reflexive and symmetric) models below.
    \begin{ctabular}{@{}c@{\;}c@{\qquad}c@{\;}c@{}}
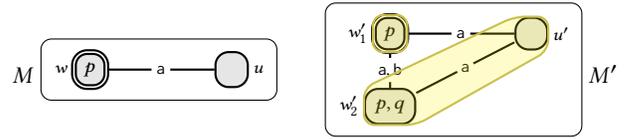
      \begin{tabular}{@{}c@{}}
        $M$
      \end{tabular}
      &
      \begin{tabular}{@{}c@{}}
        \begin{tikzpicture}[frame rectangle, node distance = 1.5em and 4.5em]
          \node [myworld, label = {[myworld-label]left:$w$}, double] (w0) {$p$};
          \node [myworld, label = {[myworld-label]right:$u$}, right = of w0] (u1) {\phantom{$p$}};

          \path (w0) edge [myarrow, -] node [myarrow-label, within] {$\aga$} (u1);
        \end{tikzpicture}
      \end{tabular}
      &
      \begin{tabular}{@{}c@{}}
        \begin{tikzpicture}[frame rectangle, node distance = 1.5em and 4.5em]
          \node [myworld, label = {[myworld-label]left:$w'_1$}, double] (w0) {$p$};
          \node [myworld, label = {[myworld-label]left:$w'_2$}, below = of w0] (u0) {$p,q$};
          \node [myworld, label = {[myworld-label]right:$u'$}, right = of w0] (u1) {};

          \path (w0) edge [myarrow, -] node [myarrow-label, within] {$\aga, \agb$} (u0)
                     edge [myarrow, -] node [myarrow-label, within] {$\aga$} (u1)
                (u0) edge [myarrow, -] node [myarrow-label, within] {$\aga$} (u1);

          \draw[topicclass] (w0.north west) -- (w0.north east) -- (w0.south east) -- (w0.south west) -- cycle;
          \draw[topicclass] (u1.north west) -- (u1.north east) -- (u1.south east) -- (u0.south east) -- (u0.south west) -- (u0.north west) -- cycle;
        \end{tikzpicture}
      \end{tabular}
      &
      \begin{tabular}{@{}c@{}}
        $M'$
      \end{tabular}
    \end{ctabular}
    Note how $(M,w) \not\Vdash \mmdiaSAE{\set{\aga, \agb}}{(\mmKa{p} \land \lnot \mmKa{\mmKa{p}})}$: making $\mmKa{p} \land \lnot \mmKa{\mmKa{p}}$ true at $w$ requires deleting the symmetric $\aga$-edge between $w$ and $u$ (so $\mmKa{p}$ holds), but this makes $u$ inaccessible for $\aga$ from $w$ (thus $\lnot \mmKa{\mmKa{p}}$ fails). Yet, $(M',w'_1) \Vdash \mmdiaSAE{\set{\aga, \agb}}{(\mmKa{p} \land \lnot \mmKa{\mmKa{p}})}$: a `conversation' among $\set{\aga, \agb}$ about $p \leftrightarrow q$ produces the desired result (see Fact \ref{fct:no.pa}). 
    
    To show that $(M,w)$ and $(M',w'_1)$ cannot be distinguished by a $q$-less formula $\alpha$ in $\LAdAPA$, \replace{proceed by}{use} structural induction over $\alpha$ and submodels of $M$ and $M'$. Both models are collectively $\pa \setminus \setn{q}$-bisimilar (witness: $\setn{(w, w'_1), (w, w'_2), (u, u')}$), so the case for atoms is immediate. As an induction hypothesis, we state that the current submodels of $M$ and $M'$ are collectively $\pa \setminus \setn{q}$-bisimilar. Boolean, epistemic, and public announcement cases follow from \autoref{thm:cobisim.LAd}. Finally, for $\mmPA{\ast}{}$ observe that for each announcement in one submodel we can always find a corresponding announcement in the other submodel such that the resulting updated models are collectively $\pa \setminus \setn{q}$-bisimilar.  This is due to the fact that each state in both models is uniquely defined by a Boolean formula containing only atoms $p$ and $q$.  Moreover, all possible updates of $\pa \setminus \setn{q}$-bisimilar submodels are given by the aforementioned witness: $\setn{(w, w'_1), (w, w'_2), (u, u')}$. E.g. if a submodel of $M'$ contains only states $w_1'$ and $w_2'$, then the corresponding submodel of $M$ would contain only state $w$.

    \smallskip

    To show $\LAdAPA \not\preccurlyeq \LAdSAE{}$, proceed in a similar fashion: consider $\mmdiaPA{\ast}{(\mmKb{p} \land \lnot \mmKb{\mmKb{p}})}$ in \LAdAPA and assume there is an equivalent $\beta \in \LAdSAE{}$. Let $q$ be an atom not occurring in $\beta$, and consider the (reflexive and symmetric) models below.
    \begin{ctabular}{@{}c@{\;}c@{\quad}c@{\;}c@{}}
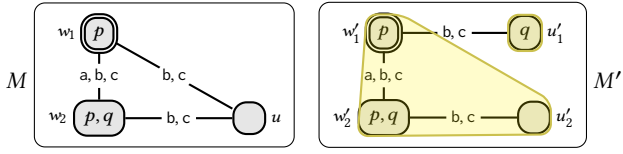
      \begin{tabular}{@{}c@{}}
        $M$
      \end{tabular}
      &
      \begin{tabular}{@{}c@{}}
        \begin{tikzpicture}[frame rectangle, node distance = 2em and 4.5em]
          \node [myworld, label = {[myworld-label]left:$w_1$}, double] (w0) {$p$};
          \node [myworld, label = {[myworld-label]left:$w_2$}, below = of w0] (w1) {$p,q$};
          \node [myworld, label = {[myworld-label]right:$u$}, right = of w1] (u) {};

          \path (w0) edge [myarrow, -] node [myarrow-label, within] {$\agb, \agc$} (u)
                     edge [myarrow, -] node [myarrow-label, within] {$\aga,\agb, \agc$} (w1)
                (w1) edge [myarrow, -] node [myarrow-label, within] {$\agb,\agc$} (u);
        \end{tikzpicture}
      \end{tabular}
      &
      \begin{tabular}{@{}c@{}}
        \begin{tikzpicture}[frame rectangle, node distance = 2em and 4.5em]
          \node [myworld, label = {[myworld-label]left:$w'_1$}, double] (w0) {$p$};
          \node [myworld, label = {[myworld-label]left:$w'_2$}, below = of w0] (w1) {$p,q$};
          \node [myworld, label = {[myworld-label]right:$u'_1$}, right = of w0] (u0) {$q$};
          \node [myworld, label = {[myworld-label]right:$u'_2$}, right = of w1] (u1) {};

          \path (w0) edge [myarrow, -] node [myarrow-label, within] {$\agb,\agc$} (u0)
                     edge [myarrow, -] node [myarrow-label, within] {$\aga,\agb, \agc$} (w1)
                (w1) edge [myarrow, -] node [myarrow-label, within] {$\agb,\agc$} (u1);

          \draw[topicclass] (w0.north west) -- (w0.north east) -- (u1.north east) -- (u1.south east) -- (w1.south west) -- cycle;
          \draw[topicclass] (u0.north west) -- (u0.north east) -- (u0.south east) -- (u0.south west) -- cycle;
        \end{tikzpicture}
      \end{tabular}
      &
      \begin{tabular}{@{}c@{}}
        $M'$
      \end{tabular}
    \end{ctabular}
    Note how $(M,w_1) \not \Vdash \mmdiaPA{\ast}{(\mmKb{p} \land \lnot \mmKb{\mmKb{p}})}$ (an announcement preserves transitivity). Yet, $(M',w'_1) \Vdash \mmdiaPA{\ast}{(\mmKb{p} \land \lnot \mmKb{\mmKb{p}})}$: the announcement of $q \rightarrow p$ (equivalence classes highlighted) produces the desired result. To show that $(M,w_1)$ and $(M',w'_1)$ cannot be distinguished by a $q$-less formula in \LAdSAE{}, \replace{proceed by}{use} structural induction. For $\mmdiaSAE{\sen}{}$, observe that the pointed models are collectively $\pa \setminus \setn{q}$-bisimilar (witness: $\setn{(w_1, w'_1), (w_2, w'_2), (u, u'_1), (u, u'_2)}$) and that, for each update in one model, there is an update in the other \replace{with the results remaining collectively $\pa \setminus \setn{q}$-bisimilar}{that preserves collective $\pa \setminus \setn{q}$-bisimilarity}. As in the previous case,  each state is uniquely characterised by a Boolean formula containing only atoms $p$ and $q$. This allows us to consider all possible bipartitions of the models\replace{. Moreover,}{, and} the witness helps to build a corresponding model. E.g., if there is a relation between $w_1'$ and $u_1'$, then we need to preserve the same relation between $w_1$ and $u$. 
  \end{proof}
\end{theorem}

\section{Discussion}\label{sec:discussion}

This paper studies further the partial communication framework of \cite{Velazquez2022}. As such, it makes sense to argue, albeit briefly, for the use of this setting as well as that of its introduced extension.

A first concern might be that, although communication \emph{between agents} is a crucial form of interaction, the public announcement logic (\ti{PAL}) framework has been already used for modelling it  (e.g., \cite{agotnes10,vanDitmarschLying2013}). Here we argue that this strategy might not be fully suited. A \ti{PAL} announcement actually requires two parameters: the announcement's precondition and the information the agents receive. When this announcement is understood as information coming from an external source, it is clear what these two parameters are, and it is clear they are the same: in order to be `announced', $\xi$ must be true, and after the announcement the agents learn that $\xi$ is the case.\footnote{More precisely, they learn $\xi$ was the case immediately before its announcement.} But when this setting is used for communication between agents, precondition and information content are not straightforward, and they might differ. When \emph{an agent} $\agi$ announces $\xi$, what is the precondition? It cannot be only $\xi$; is it enough that the agent knows $\xi$ (i.e., $\mmKi{\xi}$), or should she be introspective about it (i.e., $\mmKi{\mmKi{\xi}}$)? Analogously, what is what the other agents learn? They learn not only that $\xi$ is true; do they learn that the agent knows $\xi$ (i.e., $\mmKi{\xi}$), or even that she knows that she knows $\xi$ (i.e., $\mmKi{\mmKi{\xi}}$)? 

These questions naturally extend to situations of group communication. In group announcement logic \cite{agotnes10}, an announcement from a group $\sen$ is represented by the public announcement of $\bigwedge_{\agi \in \sen} \mmKi{\xi_{\agi}}$: each agent $\agi \in \sen$ announces, in parallel with the others, a formula she knows. However, other readings may be more appropriate: the group might announce something that is common knowledge among its members, or even announce something they all know distributively. These alternative readings are more naturally represented by the actions introduced in \cite{Baltag2010slides,AgotnesWang2017,BaltagSmets2020}, of which partial communication is a novel variation.


Then, in the partial communication setting, although only some of the agents share, this information is received by every agent in the system. One might be interested in more complex `private communication' scenarios, as those in which only some agents receive the shared information (cf., e.g., \cite{BaltagSmets2020}). Still, this `everybody hears' setting is useful for modelling classroom or meeting-like scenarios in which everybody `hears' but only some get to `talk', or for situations in which the communication channel is insecure, and thus privacy cannot be assumed. Instead of looking at extensions for modelling private communication, this paper has rather focused on the strategic aspects that arise in competitive situations. In such cases, one wonders whether there is a form of partial communication that can achieve a given goal (e.g., \cite{Ditmarsch03}). The arbitrary partial communication of \autoref{sec:SAE} can help to answer such questions.

\section{Summary and further work}\label{sec:end}

The focus of this paper is the action of \emph{partial communication}. Through it, a group of agents $\sen$ share, with every agent in the model, all the information they have about the truth-value of a formula $\chi$. Semantically, this is represented by an operation through which the uncertainty of each agent is reduced by removing the uncertainty \emph{about $\chi$} some agent in $\sen$ has already ruled out. 
After \replace{having recalled}{recalling} the \replace{framework for partial communication}{basics of this framework}, we proved that its language \LAdSSE is invariant under collective bisimulation\replace{. Moreover, we investigated}{, showing also that} the complexity of its model checking problem\replace{, and demonstrated that it}{} remains in \textit{P}\replace{}{,} as standard epistemic logic \cite{halpern92}.  It has been also shown that, while the expressivity of \LAdSSE is exactly that of the language for public announcements (both reducible to \LAd), their `update expressive power' are incomparable. The focus has then shifted to a modal operator that quantifies over the topic of the communication: a setting for \emph{arbitrary} partial communication. We have provided the operator's semantic interpretation as well as an axiom system and invariance results for the resulting language \LAdSAE{}. We have also proved that the model checking problem for the new language \LAdSAE{} is \ti{PSPACE}-complete,  similar to \ti{DELs} with action models \cite{aucher13,vandepol21} and logics with quantification over information change \cite{BalbianiBDHHL08,agotnes10,vanditmarsch17,alechina21}. Finally, we \replace{demonstrated}{showed} that \LAdSAE{} is, expressivity-wise, incomparable to the language of \emph{arbitrary} public announcements.


The framework for partial communication provides, arguably, a natural representation of communication between agents. Indeed, it works directly with the information (i.e., uncertainty) the agents have, instead of looking for formulas that are known by the agents, and then using them as announcements (as done, e.g., when dealing with group announcements \cite{agotnes10}). Additionally, the results show that this action is a truly novel epistemic action, different from others as public announcements.


There is still further work to do. In the current version of the setting, some questions still need an answer. An important one is that collective bisimulation is not `well-behaved': a model and its collective bisimulation contraction are not collectively bisimilar \cite{Roelofsen2005}. One then wonders whether there is a more adequate notion of structural equivalence for the basic language \LAd and its extensions. Then, with the partial communication setting already compared with that for public announcements (in both their basic and their `arbitrary` versions), one would like to compare it also with the setting for group announcements \cite{agotnes10}, and even with those for more general edge-removing operations (e.g., the arrow update setting \cite{KooiR11}). Finally, one can expand the presented framework. For example, one can extend the languages used here by adding a \emph{common knowledge} operator, a step that requires technical further tools \cite{AgotnesWang2017,BaltagSmets2020,GalimullinA21}. Equally interesting is a generalisation in which the topic of conversation is rather a set of formulas, together with its connection with other forms of communication (e.g., one in which some agents share \emph{all they know} with everybody).



\bibliographystyle{acm} 
\bibliography{arb-par-com}


\begin{thebibliography}{40}


\ifx \showCODEN    \undefined \def \showCODEN     #1{\unskip}     \fi
\ifx \showDOI      \undefined \def \showDOI       #1{#1}\fi
\ifx \showISBNx    \undefined \def \showISBNx     #1{\unskip}     \fi
\ifx \showISBNxiii \undefined \def \showISBNxiii  #1{\unskip}     \fi
\ifx \showISSN     \undefined \def \showISSN      #1{\unskip}     \fi
\ifx \showLCCN     \undefined \def \showLCCN      #1{\unskip}     \fi
\ifx \shownote     \undefined \def \shownote      #1{#1}          \fi
\ifx \showarticletitle \undefined \def \showarticletitle #1{#1}   \fi
\ifx \showURL      \undefined \def \showURL       {\relax}        \fi
\providecommand\bibfield[2]{#2}
\providecommand\bibinfo[2]{#2}
\providecommand\natexlab[1]{#1}
\providecommand\showeprint[2][]{arXiv:#2}

\bibitem[\protect\citeauthoryear{{\AA}gotnes, Alechina, and
  Galimullin}{{\AA}gotnes et~al\mbox{.}}{2022}]%
        {agotnes22}
\bibfield{author}{\bibinfo{person}{Thomas {\AA}gotnes},
  \bibinfo{person}{Natasha Alechina}, {and} \bibinfo{person}{Rustam
  Galimullin}.} \bibinfo{year}{2022}\natexlab{}.
\newblock \showarticletitle{Logics with Group Announcements and Distributed
  Knowledge: Completeness and Expressive Power}.
\newblock \bibinfo{journal}{\emph{Journal of Logic, Language and Information}}
  \bibinfo{volume}{31}, \bibinfo{number}{2} (\bibinfo{year}{2022}),
  \bibinfo{pages}{141--166}.
\newblock
\urldef\tempurl%
\url{https://doi.org/10.1007/s10849-022-09355-0}
\showDOI{\tempurl}


\bibitem[\protect\citeauthoryear{{\AA}gotnes, Balbiani, van Ditmarsch, and
  Seban}{{\AA}gotnes et~al\mbox{.}}{2010}]%
        {agotnes10}
\bibfield{author}{\bibinfo{person}{Thomas {\AA}gotnes},
  \bibinfo{person}{Philippe Balbiani}, \bibinfo{person}{Hans van Ditmarsch},
  {and} \bibinfo{person}{Pablo Seban}.} \bibinfo{year}{2010}\natexlab{}.
\newblock \showarticletitle{Group announcement logic}.
\newblock \bibinfo{journal}{\emph{Journal of Applied Logic}}
  \bibinfo{volume}{8}, \bibinfo{number}{1} (\bibinfo{year}{2010}),
  \bibinfo{pages}{62--81}.
\newblock
\urldef\tempurl%
\url{https://doi.org/10.1016/j.jal.2008.12.002}
\showDOI{\tempurl}


\bibitem[\protect\citeauthoryear{{\AA}gotnes and W{\'{a}}ng}{{\AA}gotnes and
  W{\'{a}}ng}{2017}]%
        {AgotnesWang2017}
\bibfield{author}{\bibinfo{person}{Thomas {\AA}gotnes} {and}
  \bibinfo{person}{Y{\`{\i}}~N. W{\'{a}}ng}.} \bibinfo{year}{2017}\natexlab{}.
\newblock \showarticletitle{Resolving distributed knowledge}.
\newblock \bibinfo{journal}{\emph{Artificial Intelligence}}
  \bibinfo{volume}{252} (\bibinfo{year}{2017}), \bibinfo{pages}{1--21}.
\newblock
\urldef\tempurl%
\url{https://doi.org/10.1016/j.artint.2017.07.002}
\showDOI{\tempurl}


\bibitem[\protect\citeauthoryear{Alechina, van Ditmarsch, Galimullin, and
  Wang}{Alechina et~al\mbox{.}}{2021}]%
        {alechina21}
\bibfield{author}{\bibinfo{person}{Natasha Alechina}, \bibinfo{person}{Hans van
  Ditmarsch}, \bibinfo{person}{Rustam Galimullin}, {and} \bibinfo{person}{Tuo
  Wang}.} \bibinfo{year}{2021}\natexlab{}.
\newblock \showarticletitle{Verification and Strategy Synthesis for Coalition
  Announcement Logic}.
\newblock \bibinfo{journal}{\emph{Journal of Logic, Language and Information}}
  \bibinfo{volume}{30}, \bibinfo{number}{4} (\bibinfo{year}{2021}),
  \bibinfo{pages}{671--700}.
\newblock
\urldef\tempurl%
\url{https://doi.org/10.1007/s10849-021-09339-6}
\showDOI{\tempurl}


\bibitem[\protect\citeauthoryear{Aucher and Schwarzentruber}{Aucher and
  Schwarzentruber}{2013}]%
        {aucher13}
\bibfield{author}{\bibinfo{person}{Guillaume Aucher} {and}
  \bibinfo{person}{Fran{\c{c}}ois Schwarzentruber}.}
  \bibinfo{year}{2013}\natexlab{}.
\newblock \showarticletitle{On the Complexity of Dynamic Epistemic Logic}. In
  \bibinfo{booktitle}{\emph{Proceedings of the 14th {TARK}}},
  \bibfield{editor}{\bibinfo{person}{Burkhard~C. Schipper}} (Ed.).
\newblock


\bibitem[\protect\citeauthoryear{Balbiani, Baltag, van Ditmarsch, Herzig,
  Hoshi, and de~Lima}{Balbiani et~al\mbox{.}}{2008}]%
        {BalbianiBDHHL08}
\bibfield{author}{\bibinfo{person}{Philippe Balbiani},
  \bibinfo{person}{Alexandru Baltag}, \bibinfo{person}{Hans van Ditmarsch},
  \bibinfo{person}{Andreas Herzig}, \bibinfo{person}{Tomohiro Hoshi}, {and}
  \bibinfo{person}{Tiago de Lima}.} \bibinfo{year}{2008}\natexlab{}.
\newblock \showarticletitle{'Knowable' as 'known after an announcement'}.
\newblock \bibinfo{journal}{\emph{The Review of Symbolic Logic}}
  \bibinfo{volume}{1}, \bibinfo{number}{3} (\bibinfo{year}{2008}),
  \bibinfo{pages}{305--334}.
\newblock
\urldef\tempurl%
\url{https://doi.org/10.1017/S1755020308080210}
\showDOI{\tempurl}


\bibitem[\protect\citeauthoryear{Balbiani and van Ditmarsch}{Balbiani and van
  Ditmarsch}{2015}]%
        {balbiani15}
\bibfield{author}{\bibinfo{person}{Philippe Balbiani} {and}
  \bibinfo{person}{Hans van Ditmarsch}.} \bibinfo{year}{2015}\natexlab{}.
\newblock \showarticletitle{A simple proof of the completeness of {APAL}}.
\newblock \bibinfo{journal}{\emph{Studies in Logic}} \bibinfo{volume}{8},
  \bibinfo{number}{2} (\bibinfo{year}{2015}), \bibinfo{pages}{65--78}.
\newblock


\bibitem[\protect\citeauthoryear{Baltag}{Baltag}{2010}]%
        {Baltag2010slides}
\bibfield{author}{\bibinfo{person}{Alexandru Baltag}.}
  \bibinfo{year}{2010}\natexlab{}.
\newblock \bibinfo{title}{What is DEL good for?}  (\bibinfo{year}{2010}).
\newblock
\urldef\tempurl%
\url{http://ai.stanford.edu/~epacuit/lograt/esslli2010-slides/copenhagenesslli.pdf}
\showURL{%
\tempurl}
\newblock
\shownote{{Workshop on Logic, Rationality and Intelligent Interaction}.}


\bibitem[\protect\citeauthoryear{Baltag, Moss, and Solecki}{Baltag
  et~al\mbox{.}}{1998}]%
        {BaltagMS98}
\bibfield{author}{\bibinfo{person}{Alexandru Baltag},
  \bibinfo{person}{Lawrence~S. Moss}, {and} \bibinfo{person}{S{\l}awomir
  Solecki}.} \bibinfo{year}{1998}\natexlab{}.
\newblock \showarticletitle{The Logic of Public Announcements and Common
  Knowledge and Private Suspicions}. In \bibinfo{booktitle}{\emph{Proceedings
  of the 7th TARK}}, \bibfield{editor}{\bibinfo{person}{Itzhak Gilboa}} (Ed.).
  \bibinfo{publisher}{Morgan Kaufmann}, \bibinfo{pages}{43--56}.
\newblock
\showISBNx{1-55860-563-0}


\bibitem[\protect\citeauthoryear{Baltag and Smets}{Baltag and Smets}{2020}]%
        {BaltagSmets2020}
\bibfield{author}{\bibinfo{person}{Alexandru Baltag} {and}
  \bibinfo{person}{Sonja Smets}.} \bibinfo{year}{2020}\natexlab{}.
\newblock \showarticletitle{Learning What Others Know}. In
  \bibinfo{booktitle}{\emph{{LPAR} 2020}} \emph{(\bibinfo{series}{EPiC Series
  in Computing}, Vol.~\bibinfo{volume}{73})},
  \bibfield{editor}{\bibinfo{person}{Elvira Albert} {and}
  \bibinfo{person}{Laura Kov{\'{a}}cs}} (Eds.). \bibinfo{publisher}{EasyChair},
  \bibinfo{pages}{90--119}.
\newblock
\urldef\tempurl%
\url{https://doi.org/10.29007/plm4}
\showDOI{\tempurl}


\bibitem[\protect\citeauthoryear{Baltag and Smets}{Baltag and Smets}{2021}]%
        {BaltagSmets2021}
\bibfield{author}{\bibinfo{person}{Alexandru Baltag} {and}
  \bibinfo{person}{Sonja Smets}.} \bibinfo{year}{2021}\natexlab{}.
\newblock \showarticletitle{Learning What Others Know}.
\newblock \bibinfo{journal}{\emph{CoRR}}  \bibinfo{volume}{abs/2109.07255}
  (\bibinfo{year}{2021}).
\newblock
\showeprint[arXiv]{2109.07255}
\urldef\tempurl%
\url{https://arxiv.org/abs/2109.07255}
\showURL{%
\tempurl}


\bibitem[\protect\citeauthoryear{de~Bruin}{de~Bruin}{2010}]%
        {deBruin2010}
\bibfield{author}{\bibinfo{person}{Boudewijn de Bruin}.}
  \bibinfo{year}{2010}\natexlab{}.
\newblock \bibinfo{booktitle}{\emph{Explaining Games: The Epistemic Programme
  in Game Theory}}.
\newblock \bibinfo{publisher}{Springer}, \bibinfo{address}{Dordrecht}.
\newblock
\showISBNx{978-1-4020-9905-2}
\urldef\tempurl%
\url{https://doi.org/10.1007/978-1-4020-9906-9}
\showDOI{\tempurl}


\bibitem[\protect\citeauthoryear{de~Haan and van~de Pol}{de~Haan and van~de
  Pol}{2021}]%
        {vandepol21}
\bibfield{author}{\bibinfo{person}{Ronald de Haan} {and} \bibinfo{person}{Iris
  van~de Pol}.} \bibinfo{year}{2021}\natexlab{}.
\newblock \showarticletitle{On the Computational Complexity of Model Checking
  for Dynamic Epistemic Logic with {S5} Models}.
\newblock \bibinfo{journal}{\emph{{FLAP}}} \bibinfo{volume}{8},
  \bibinfo{number}{3} (\bibinfo{year}{2021}), \bibinfo{pages}{621--658}.
\newblock


\bibitem[\protect\citeauthoryear{Fagin, Halpern, Moses, and Vardi}{Fagin
  et~al\mbox{.}}{1995}]%
        {FaginHalpernMosesVardi1995}
\bibfield{author}{\bibinfo{person}{Ronald Fagin}, \bibinfo{person}{Joseph~Y.
  Halpern}, \bibinfo{person}{Yoram Moses}, {and} \bibinfo{person}{Moshe~Y.
  Vardi}.} \bibinfo{year}{1995}\natexlab{}.
\newblock \bibinfo{booktitle}{\emph{Reasoning about knowledge}}.
\newblock \bibinfo{publisher}{The MIT Press}, \bibinfo{address}{Cambridge,
  Mass.}
\newblock
\showISBNx{0-262-06162-7}


\bibitem[\protect\citeauthoryear{Galimullin and {\AA}gotnes}{Galimullin and
  {\AA}gotnes}{2021}]%
        {GalimullinA21}
\bibfield{author}{\bibinfo{person}{Rustam Galimullin} {and}
  \bibinfo{person}{Thomas {\AA}gotnes}.} \bibinfo{year}{2021}\natexlab{}.
\newblock \showarticletitle{Quantified Announcements and Common Knowledge}. In
  \bibinfo{booktitle}{\emph{Proceegins of the 20th {AAMAS}}},
  \bibfield{editor}{\bibinfo{person}{Frank Dignum}, \bibinfo{person}{Alessio
  Lomuscio}, \bibinfo{person}{Ulle Endriss}, {and} \bibinfo{person}{Ann
  Now{\'{e}}}} (Eds.). \bibinfo{publisher}{{ACM}}, \bibinfo{pages}{528--536}.
\newblock
\urldef\tempurl%
\url{https://dl.acm.org/doi/10.5555/3463952.3464018}
\showURL{%
\tempurl}


\bibitem[\protect\citeauthoryear{Gerbrandy and Groeneveld}{Gerbrandy and
  Groeneveld}{1997}]%
        {GerbrandyGroeneveld1997}
\bibfield{author}{\bibinfo{person}{Jelle Gerbrandy} {and}
  \bibinfo{person}{Willem Groeneveld}.} \bibinfo{year}{1997}\natexlab{}.
\newblock \showarticletitle{Reasoning about information change}.
\newblock \bibinfo{journal}{\emph{Journal of Logic, Language, and Information}}
  \bibinfo{volume}{6}, \bibinfo{number}{2} (\bibinfo{year}{1997}),
  \bibinfo{pages}{147--196}.
\newblock
\urldef\tempurl%
\url{https://doi.org/10.1023/A:1008222603071}
\showDOI{\tempurl}


\bibitem[\protect\citeauthoryear{Goldblatt}{Goldblatt}{1982}]%
        {goldblatt82}
\bibfield{author}{\bibinfo{person}{Robert Goldblatt}.}
  \bibinfo{year}{1982}\natexlab{}.
\newblock \bibinfo{booktitle}{\emph{Axiomatising the Logic of Computer
  Programming}}. \bibinfo{series}{LNCS}, Vol.~\bibinfo{volume}{130}.
\newblock \bibinfo{publisher}{Springer}.
\newblock
\urldef\tempurl%
\url{https://doi.org/10.1007/BFb0022481}
\showDOI{\tempurl}


\bibitem[\protect\citeauthoryear{Halpern and Moses}{Halpern and Moses}{1990}]%
        {HalpernM90}
\bibfield{author}{\bibinfo{person}{Joseph~Y. Halpern} {and}
  \bibinfo{person}{Yoram Moses}.} \bibinfo{year}{1990}\natexlab{}.
\newblock \showarticletitle{Knowledge and Common Knowledge in a Distributed
  Environment}.
\newblock \bibinfo{journal}{\emph{Journal of the {ACM}}} \bibinfo{volume}{37},
  \bibinfo{number}{3} (\bibinfo{year}{1990}), \bibinfo{pages}{549--587}.
\newblock
\urldef\tempurl%
\url{https://doi.org/10.1145/79147.79161}
\showDOI{\tempurl}


\bibitem[\protect\citeauthoryear{Halpern and Moses}{Halpern and Moses}{1992}]%
        {halpern92}
\bibfield{author}{\bibinfo{person}{Joseph~Y. Halpern} {and}
  \bibinfo{person}{Yoram Moses}.} \bibinfo{year}{1992}\natexlab{}.
\newblock \showarticletitle{A Guide to Completeness and Complexity for Modal
  Logics of Knowledge and Belief}.
\newblock \bibinfo{journal}{\emph{Artificial Intelligence}}
  \bibinfo{volume}{54}, \bibinfo{number}{2} (\bibinfo{year}{1992}),
  \bibinfo{pages}{319--379}.
\newblock
\urldef\tempurl%
\url{https://doi.org/10.1016/0004-3702(92)90049-4}
\showDOI{\tempurl}


\bibitem[\protect\citeauthoryear{Hendricks}{Hendricks}{2006}]%
        {PhilStu:EpisLog}
\bibfield{editor}{\bibinfo{person}{Vincent~F. Hendricks}} (Ed.).
  \bibinfo{year}{2006}\natexlab{}.
\newblock \bibinfo{booktitle}{\emph{8 Bridges between Formal and Mainstream
  Epistemology}}.
\newblock
\newblock
\shownote{\textit{Philosophical Studies}, 128(1).}


\bibitem[\protect\citeauthoryear{Hintikka}{Hintikka}{1962}]%
        {Hintikka1962}
\bibfield{author}{\bibinfo{person}{Jaakko Hintikka}.}
  \bibinfo{year}{1962}\natexlab{}.
\newblock \bibinfo{booktitle}{\emph{Knowledge and Belief}}.
\newblock \bibinfo{publisher}{Cornell University Press},
  \bibinfo{address}{Ithaca, N.Y.}
\newblock
\showISBNx{1-904987-08-7}


\bibitem[\protect\citeauthoryear{Kanellakis and Smolka}{Kanellakis and
  Smolka}{1990}]%
        {kanellakis90}
\bibfield{author}{\bibinfo{person}{Paris~C. Kanellakis} {and}
  \bibinfo{person}{Scott~A. Smolka}.} \bibinfo{year}{1990}\natexlab{}.
\newblock \showarticletitle{{CCS} Expressions, Finite State Processes, and
  Three Problems of Equivalence}.
\newblock \bibinfo{journal}{\emph{Information and Computation}}
  \bibinfo{volume}{86}, \bibinfo{number}{1} (\bibinfo{year}{1990}),
  \bibinfo{pages}{43--68}.
\newblock
\urldef\tempurl%
\url{https://doi.org/10.1016/0890-5401(90)90025-D}
\showDOI{\tempurl}


\bibitem[\protect\citeauthoryear{Kooi and Renne}{Kooi and Renne}{2011}]%
        {KooiR11}
\bibfield{author}{\bibinfo{person}{Barteld Kooi} {and} \bibinfo{person}{Bryan
  Renne}.} \bibinfo{year}{2011}\natexlab{}.
\newblock \showarticletitle{Arrow Update Logic}.
\newblock \bibinfo{journal}{\emph{The Review of Symbolic Logic}}
  \bibinfo{volume}{4}, \bibinfo{number}{4} (\bibinfo{year}{2011}),
  \bibinfo{pages}{536--559}.
\newblock
\urldef\tempurl%
\url{https://doi.org/10.1017/S1755020311000189}
\showDOI{\tempurl}


\bibitem[\protect\citeauthoryear{Kuijer}{Kuijer}{2015}]%
        {kuijer15}
\bibfield{author}{\bibinfo{person}{Louwe~B. Kuijer}.}
  \bibinfo{year}{2015}\natexlab{}.
\newblock \showarticletitle{An Arrow-based Dynamic Logic of Norms}. In
  \bibinfo{booktitle}{\emph{Proceedings of the 3rd {SR}}},
  \bibfield{editor}{\bibinfo{person}{Julian Gutierrez}, \bibinfo{person}{Fabio
  Mogavero}, \bibinfo{person}{Aniello Murano}, {and} \bibinfo{person}{Michael
  Wooldridge}} (Eds.). \bibinfo{pages}{1--11}.
\newblock


\bibitem[\protect\citeauthoryear{Lutz}{Lutz}{2006}]%
        {Lutz2006}
\bibfield{author}{\bibinfo{person}{Carsten Lutz}.}
  \bibinfo{year}{2006}\natexlab{}.
\newblock \showarticletitle{Complexity and succinctness of public announcement
  logic}. In \bibinfo{booktitle}{\emph{Proceedings of the 5th {AAMAS}}},
  \bibfield{editor}{\bibinfo{person}{Hideyuki Nakashima},
  \bibinfo{person}{Michael~P. Wellman}, \bibinfo{person}{Gerhard Weiss}, {and}
  \bibinfo{person}{Peter Stone}} (Eds.). \bibinfo{publisher}{{ACM}},
  \bibinfo{pages}{137--143}.
\newblock
\showISBNx{1-59593-303-4}
\urldef\tempurl%
\url{https://doi.org/10.1145/1160633.1160657}
\showDOI{\tempurl}


\bibitem[\protect\citeauthoryear{Meyer and van~der Hoek}{Meyer and van~der
  Hoek}{1995}]%
        {MeyervanDerHoek1995elaics}
\bibfield{author}{\bibinfo{person}{John-Jules~Ch. Meyer} {and}
  \bibinfo{person}{Wiebe van~der Hoek}.} \bibinfo{year}{1995}\natexlab{}.
\newblock \bibinfo{booktitle}{\emph{Epistemic Logic for {AI} and Computer
  Science}}.
\newblock \bibinfo{publisher}{CUP}.
\newblock
\showISBNx{0-521-46014-7}
\urldef\tempurl%
\url{https://doi.org/10.1017/CBO9780511569852}
\showDOI{\tempurl}


\bibitem[\protect\citeauthoryear{Plaza}{Plaza}{1989}]%
        {Plaza1989}
\bibfield{author}{\bibinfo{person}{Jan~A. Plaza}.}
  \bibinfo{year}{1989}\natexlab{}.
\newblock \showarticletitle{Logics of public communications}. In
  \bibinfo{booktitle}{\emph{Proceedings of the 4th International Symposium on
  Methodologies for Intelligent Systems}},
  \bibfield{editor}{\bibinfo{person}{M.~L. Emrich}, \bibinfo{person}{M.~S.
  Pfeifer}, \bibinfo{person}{M.~Hadzikadic}, {and} \bibinfo{person}{Z.~W. Ras}}
  (Eds.). \bibinfo{pages}{201--216}.
\newblock


\bibitem[\protect\citeauthoryear{Roelofsen}{Roelofsen}{2005}]%
        {Roelofsen2005}
\bibfield{author}{\bibinfo{person}{Floris Roelofsen}.}
  \bibinfo{year}{2005}\natexlab{}.
\newblock \bibinfo{title}{Bisimulation and Distributed Knowledge Revisited}.
  (\bibinfo{year}{2005}).
\newblock
\newblock
\shownote{Available at
  \url{https://projects.illc.uva.nl/lgc/papers/d-know.pdf}.}


\bibitem[\protect\citeauthoryear{Roelofsen}{Roelofsen}{2007}]%
        {Roelofsen2007}
\bibfield{author}{\bibinfo{person}{Floris Roelofsen}.}
  \bibinfo{year}{2007}\natexlab{}.
\newblock \showarticletitle{Distributed knowledge}.
\newblock \bibinfo{journal}{\emph{Journal of Applied Non-Classical Logics}}
  \bibinfo{volume}{17}, \bibinfo{number}{2} (\bibinfo{year}{2007}),
  \bibinfo{pages}{255--273}.
\newblock
\urldef\tempurl%
\url{https://doi.org/10.3166/jancl.17.255-273}
\showDOI{\tempurl}


\bibitem[\protect\citeauthoryear{van Benthem}{van Benthem}{2011}]%
        {vanBenthem2011ldii}
\bibfield{author}{\bibinfo{person}{Johan van Benthem}.}
  \bibinfo{year}{2011}\natexlab{}.
\newblock \bibinfo{booktitle}{\emph{Logical Dynamics of Information and
  Interaction}}.
\newblock \bibinfo{publisher}{CUP}.
\newblock
\showISBNx{978-0-521-76579-4}


\bibitem[\protect\citeauthoryear{van Benthem and Liu}{van Benthem and
  Liu}{2007}]%
        {BenthemL07}
\bibfield{author}{\bibinfo{person}{Johan van Benthem} {and}
  \bibinfo{person}{Fenrong Liu}.} \bibinfo{year}{2007}\natexlab{}.
\newblock \showarticletitle{Dynamic logic of preference upgrade}.
\newblock \bibinfo{journal}{\emph{Journal of Applied Non-Classical Logics}}
  \bibinfo{volume}{17}, \bibinfo{number}{2} (\bibinfo{year}{2007}),
  \bibinfo{pages}{157--182}.
\newblock
\urldef\tempurl%
\url{https://doi.org/10.3166/jancl.17.157-182}
\showDOI{\tempurl}


\bibitem[\protect\citeauthoryear{van Ditmarsch}{van Ditmarsch}{2003}]%
        {Ditmarsch03}
\bibfield{author}{\bibinfo{person}{Hans van Ditmarsch}.}
  \bibinfo{year}{2003}\natexlab{}.
\newblock \showarticletitle{The Russian Cards Problem}.
\newblock \bibinfo{journal}{\emph{Studia Logica}} \bibinfo{volume}{75},
  \bibinfo{number}{1} (\bibinfo{year}{2003}), \bibinfo{pages}{31--62}.
\newblock
\urldef\tempurl%
\url{https://doi.org/10.1023/A:1026168632319}
\showDOI{\tempurl}


\bibitem[\protect\citeauthoryear{van Ditmarsch}{van Ditmarsch}{2014}]%
        {vanDitmarschLying2013}
\bibfield{author}{\bibinfo{person}{Hans van Ditmarsch}.}
  \bibinfo{year}{2014}\natexlab{}.
\newblock \showarticletitle{Dynamics of lying}.
\newblock \bibinfo{journal}{\emph{Synthese}} \bibinfo{volume}{191},
  \bibinfo{number}{5} (\bibinfo{year}{2014}), \bibinfo{pages}{745--777}.
\newblock
\showISSN{0039-7857}
\urldef\tempurl%
\url{https://doi.org/10.1007/s11229-013-0275-3}
\showDOI{\tempurl}


\bibitem[\protect\citeauthoryear{van Ditmarsch}{van Ditmarsch}{2020}]%
        {vanditmarsch20}
\bibfield{author}{\bibinfo{person}{Hans van Ditmarsch}.}
  \bibinfo{year}{2020}\natexlab{}.
\newblock \showarticletitle{To Be Announced}.
\newblock \bibinfo{journal}{\emph{CoRR}}  \bibinfo{volume}{abs/2004.05802}
  (\bibinfo{year}{2020}).
\newblock
\showeprint[arXiv]{2004.05802}
\urldef\tempurl%
\url{https://arxiv.org/abs/2004.05802}
\showURL{%
\tempurl}


\bibitem[\protect\citeauthoryear{van Ditmarsch, Fern{\'{a}}ndez{-}Duque, and
  van~der Hoek}{van Ditmarsch et~al\mbox{.}}{2014}]%
        {vanditmarsch14}
\bibfield{author}{\bibinfo{person}{Hans van Ditmarsch}, \bibinfo{person}{David
  Fern{\'{a}}ndez{-}Duque}, {and} \bibinfo{person}{Wiebe van~der Hoek}.}
  \bibinfo{year}{2014}\natexlab{}.
\newblock \showarticletitle{On the definability of simulation and bisimulation
  in epistemic logic}.
\newblock \bibinfo{journal}{\emph{Journal of Logic and Computation}}
  \bibinfo{volume}{24}, \bibinfo{number}{6} (\bibinfo{year}{2014}),
  \bibinfo{pages}{1209--1227}.
\newblock
\urldef\tempurl%
\url{https://doi.org/10.1093/logcom/exs058}
\showDOI{\tempurl}


\bibitem[\protect\citeauthoryear{van Ditmarsch, van~der Hoek, and Kooi}{van
  Ditmarsch et~al\mbox{.}}{2008}]%
        {vanDitmarschEtAl2007}
\bibfield{author}{\bibinfo{person}{Hans van Ditmarsch}, \bibinfo{person}{Wiebe
  van~der Hoek}, {and} \bibinfo{person}{Barteld Kooi}.}
  \bibinfo{year}{2008}\natexlab{}.
\newblock \bibinfo{booktitle}{\emph{Dynamic Epistemic Logic}}.
\newblock \bibinfo{publisher}{Springer}, \bibinfo{address}{Dordrecht, The
  Netherlands}.
\newblock
\showISBNx{978-1-4020-5838-7}
\urldef\tempurl%
\url{https://doi.org/10.1007/978-1-4020-5839-4}
\showDOI{\tempurl}


\bibitem[\protect\citeauthoryear{van Ditmarsch, van~der Hoek, Kooi, and
  Kuijer}{van Ditmarsch et~al\mbox{.}}{2017}]%
        {vanditmarsch17}
\bibfield{author}{\bibinfo{person}{Hans van Ditmarsch}, \bibinfo{person}{Wiebe
  van~der Hoek}, \bibinfo{person}{Barteld Kooi}, {and}
  \bibinfo{person}{Louwe~B. Kuijer}.} \bibinfo{year}{2017}\natexlab{}.
\newblock \showarticletitle{Arbitrary arrow update logic}.
\newblock \bibinfo{journal}{\emph{Artificial Intelligence}}
  \bibinfo{volume}{242} (\bibinfo{year}{2017}), \bibinfo{pages}{80--106}.
\newblock
\urldef\tempurl%
\url{https://doi.org/10.1016/j.artint.2016.10.003}
\showDOI{\tempurl}


\bibitem[\protect\citeauthoryear{Vel{\'{a}}zquez{-}Quesada}{Vel{\'{a}}zquez{-}Quesada}{2022}]%
        {Velazquez2022}
\bibfield{author}{\bibinfo{person}{Fernando~R. Vel{\'{a}}zquez{-}Quesada}.}
  \bibinfo{year}{2022}\natexlab{}.
\newblock \showarticletitle{Communication between agents in dynamic epistemic
  logic}.
\newblock \bibinfo{journal}{\emph{CoRR}}  \bibinfo{volume}{abs/2210.04656}
  (\bibinfo{year}{2022}).
\newblock
\urldef\tempurl%
\url{https://doi.org/10.48550/arXiv.2210.04656}
\showDOI{\tempurl}
\showeprint[arXiv]{2210.04656}


\bibitem[\protect\citeauthoryear{Wang and Cao}{Wang and Cao}{2013}]%
        {WangC13}
\bibfield{author}{\bibinfo{person}{Yanjing Wang} {and}
  \bibinfo{person}{Qinxiang Cao}.} \bibinfo{year}{2013}\natexlab{}.
\newblock \showarticletitle{On axiomatizations of public announcement logic}.
\newblock \bibinfo{journal}{\emph{Synthese}} \bibinfo{volume}{190},
  \bibinfo{number}{Supplement-1} (\bibinfo{year}{2013}),
  \bibinfo{pages}{103--134}.
\newblock
\urldef\tempurl%
\url{https://doi.org/10.1007/s11229-012-0233-5}
\showDOI{\tempurl}


\bibitem[\protect\citeauthoryear{W{\'{a}}ng and {\AA}gotnes}{W{\'{a}}ng and
  {\AA}gotnes}{2013}]%
        {WangA13}
\bibfield{author}{\bibinfo{person}{Y{\`{\i}}~N. W{\'{a}}ng} {and}
  \bibinfo{person}{Thomas {\AA}gotnes}.} \bibinfo{year}{2013}\natexlab{}.
\newblock \showarticletitle{Public announcement logic with distributed
  knowledge: expressivity, completeness and complexity}.
\newblock \bibinfo{journal}{\emph{Synthese}} \bibinfo{volume}{190},
  \bibinfo{number}{Supplement-1} (\bibinfo{year}{2013}),
  \bibinfo{pages}{135--162}.
\newblock
\urldef\tempurl%
\url{https://doi.org/10.1007/s11229-012-0243-3}
\showDOI{\tempurl}


\end{thebibliography}


\clearpage

\appendix
\section{Appendix}\label{sec:appendix}

\subsection*{Proof of \autoref{thm:cobisim.LAdSSE}}\label{thm:cobisim.LAdSSE:proof}

Since \LAdSSE is the union of $\LAdSSE^i$ for all $i \in \Nat$, the proof will proceed by induction on $i$. In fact, the manuscript will prove a stronger statement: for every $\psi \in \LAdSSE$ and every $M = \tupla{W, R, V}$ and $M' = \tupla{W', R', V'}$, if $(M,w) \cobisim (M',w')$ then \begin{inlineenumco} \item\label{itm:cobisim.LAdSSE.invariant} $(M,w) \Vdash \psi$ if and only if $(M',w') \Vdash \psi$, and \item\label{itm:cobisim.LAdSSE.preserves} $(\mopSSE{M}{\sen}{\psi},w) \cobisim (\mopSSE{M'}{\sen}{\psi},w')$\end{inlineenumco}.

\ssparagraph{Base case.} Take $\psi \in \LAdSSE^0 = \LAd$. In this case, \autoref{itm:cobisim.LAdSSE.invariant} is nothing but Theorem 2.7.  For \autoref{itm:cobisim.LAdSSE.preserves}, suppose $(M,w) \cobisim (M',w')$ and let $Z$ be the witness; it will be shown that $Z$ is also a collective $\pa$-bisimulation between $\mopSSE{M}{\sen}{\psi} = \tuplan{W, \relSSE{R}{\sen}{\psi}, V}$ and $\mopSSE{M'}{\sen}{\psi} = \tuplan{W', \relSSE{R'}{\sen}{\psi}, V'}$. Take any $(u,u') \in Z$.
\begin{compactitemize}
  \item \tb{Atoms}. The operation does not change atomic valuations. Thus, since $Z$ satisfies \tb{atoms} for $M$ and $M'$, it also satisfies it for $\mopSSE{M}{\sen}{\psi}$ and $\mopSSE{M'}{\sen}{\psi}$.

  \item \tb{Forth}. Take any $\sag \subseteq \ag$ and any $v \in W$ such that $\sbDg{\relSSE{R}{\sen}{\psi}}uv$. Since $\sbDg{\relSSE{R}{\sen}{\psi}} = \sbDs{R}{\sag \cup \sen} \cup (\sbDs{R}{\sag} \cap {\knonfu{M}{\psi}})$ (Footnote 5), then $\sbDs{R}{\sag \cup \sen}uv$ or $(\sbDs{R}{\sag} \cap {\knonfu{M}{\psi}})uv$. \begin{inlineenum} \item If $\sbDs{R}{\sag \cup \sen}uv$ then, since $Z$ satisfies \tb{forth} for $M$ and $M'$, there is $v' \in W'$ such that $\sbDs{R'}{\sag \cup \sen}u'v'$ and $(v,v') \in Z$. Since $\sbDg{\relSSE{R'}{\sen}{\psi}} = \sbDs{R'}{\sag \cup \sen} \cup (\sbDs{R'}{\sag} \cap {\knonfu{M'}{\psi}})$, the former implies $\sbDg{\relSSE{R'}{\sen}{\psi}}u'v'$. Thus, there is $v' \in W'$ such that $\sbDg{\relSSE{R'}{\sen}{\psi}}u'v'$ and $(v,v') \in Z$, as required. \item If $(\sbDs{R}{\sag} \cap {\knonfu{M}{\psi}})uv$, then both $\sbDs{R}{\sag}uv$ and $u \knonfu{M}{\psi} v$. From the first and the fact that $Z$ satisfies \tb{forth} for $M$ and $M'$, there is $v' \in W'$ such that $\sbDs{R'}{\sag}u'v'$ and $(v,v') \in Z$. Now, $u \knonfu{M}{\psi} v$ indicates that $u$ and $v$ agree on $\psi$'s truth-value. But $\psi \in \LAd$. Thus, \autoref{itm:cobisim.LAdSSE.invariant} from this base case indicates that $u$ and $u'$ also agree on $\psi$ (from $(u,u') \in Z$), and so do $v$ and $v'$ (from $(v,v') \in Z$). Hence, $u'$ and $v'$ agree on $\psi$'s truth-value, that is, $u' \knonfu{M'}{\psi} v'$. Therefore, $(\sbDs{R'}{\sag} \cap {\knonfu{M'}{\psi}})uv$, so $\sbDg{\relSSE{R'}{\sen}{\psi}}u'v'$. This means there is $v' \in W'$ such that $\sbDg{\relSSE{R'}{\sen}{\psi}}u'v'$ and $(v,v') \in Z$, as required.\end{inlineenum}

  \item \tb{Back}. As in \tb{forth}, using the fact that $Z$ satisfies \tb{back} for $M$ and $M'$.
\end{compactitemize}

Thus, $\mopSSE{M}{\sen}{\psi} \cobisim \mopSSE{M'}{\sen}{\psi}$. Finally, $(w,w') \in Z$, so $(\mopSSE{M}{\sen}{\psi},w)$ $\cobisim$ $(\mopSSE{M'}{\sen}{\psi},w')$.

\ssparagraph{Inductive case.} Take $\psi \in \LAdSSE^{n+1}$ and suppose $(M,w) \cobisim (M',w')$. For \autoref{itm:cobisim.LAdSSE.invariant}, proceed by structural induction on $\psi$. The cases for atoms, Boolean operators and $\modDg$ are as in Theorem 2.7. The remaining case is for formulas of the form $\mmSSE{\sen}{\chi}{\varphi}$ with $\chi \in \LAdSSE^{n}$ and $\varphi \in \LAdSSE^{n+1}$. Here, the structural IH states that collectively $\pa$-bisimilar pointed models agree on the truth value of the subformula $\varphi$. Note how, since $\chi \in \LAdSSE^{n}$ and $(M,w) \cobisim (M',w')$, \autoref{itm:cobisim.LAdSSE.preserves} of the (global) IH implies $(\mopSSE{M}{\sen}{\chi},w) \cobisim (\mopSSE{M'}{\sen}{\chi},w')$. Now, from left to right, suppose $(M, w) \Vdash \mmSSE{\sen}{\chi}{\varphi}$. By semantic interpretation, $(\mopSSE{M}{\sen}{\chi}, w) \Vdash \varphi$; thus, from the structural IH, $(\mopSSE{M'}{\sen}{\chi}, w') \Vdash \varphi$, i.e., $(M', w') \Vdash \mmSSE{\sen}{\chi}{\varphi}$. The right-to-left direction is analogous.

It is only left to prove \autoref{itm:cobisim.LAdSSE.preserves} for $\psi \in \LAdSSE^{n+1}$. This can be done as in the (global) base case, using \autoref{itm:cobisim.LAdSSE.invariant} from this inductive case instead.

\subsection*{Proposition A.1}\label{pro:wdPA.eq.RDpa}

Let $M = \tupla{W, \R, V}$ be a model; let $\xi$ be a formula. Recall \citep{Plaza1989} that the world-removing public announcement of $\xi$ on $M$ yields the model $\mopPA{M'}{\xi} = \tuplan{\truthset{M}{\xi}, \setn{ \sbi{\R'} \mid \agi \in \ag} , V'}$ with
\[
  \sbi{\R'} := \sbi{\R} \cap (\truthset{M}{\xi} \times \truthset{M}{\xi})
  \qquad\text{and}\qquad  
  V'(p) := V(p) \cap \truthset{M}{\xi}.
\]
Now, take any $w \in \dom{\mopPA{M'}{\xi}}$. Then,
\begin{ctabular}{c}
  $(\mopPA{M}{\xi}, w) \cobisim (\mopPA{M'}{\xi}, w)$.
\end{ctabular}
\begin{proof}[Proof sketch]
  Intuitively, the difference between the world-removing and edge-deleting approaches makes no difference for a collective bisimulation: in both cases, the $\lnot\xi$-partition becomes inaccessible from the $\xi$-partition, where the world $w$ lies. Formally, it is enough to prove that the relation
  \[ Z := \setn{ (u,u) \in (W \times \truthset{M}{\xi}) \mid u \in \truthset{M}{\xi} } \]
  is a collective $\pa$-bisimulation (between $\mopPA{M}{\xi}$ and $\mopPA{M'}{\xi}$) containing the pair $(w,w)$.
\end{proof}

\subsection*{Theorem A.1}\label{thm:cobisim.LAdPA}

Let $(M,w)$ and $(M', w')$ be two pointed models. If $(M,w) \cobisim (M',w')$ then, for every $\psi \in \LAdPA$,
\begin{ctabular}{c}
  $(M,w) \Vdash \psi$ \qtnq{if and only if} $(M',w') \Vdash \psi$.
\end{ctabular}
\begin{proof}
  Analogous to the \hyperref[thm:cobisim.LAdSSE:proof]{proof} of \autoref{thm:cobisim.LAdSSE}.
\end{proof}

\subsection*{Proof of \autoref{thm:cobisim.LAdSAE}}\label{thm:cobisim.LAdSAE:proof}

Since \LAdSAE{} is the union of $\LAdSAE{,i}$ for all $i \in \Nat$, proceed again by induction on $i$ (as in the proof of Theorem 2.11). Again, one proves a stronger statement: for every $\psi \in \LAdSAE{}$ and every $M = \tupla{W, R, V}$ and $M' = \tupla{W', R', V'}$, if $(M,w) \cobisim (M',w')$ then \begin{inlineenumco} \item\label{itm:cobisim.LAdSAE.invariant} $(M,w) \Vdash \psi$ if and only if $(M',w') \Vdash \psi$, and \item\label{itm:cobisim.LAdSAE.preserves} $(\mopSSE{M}{\sen}{\psi},w) \cobisim (\mopSSE{M'}{\sen}{\psi},w')$\end{inlineenumco}.

\ssparagraph{Base case.} This base case is for formulas in $\LAdSAE{,0} = \LAd^\ast$, defined as \LAd plus the modality $\modboxSAE{\sen}$. For \autoref{itm:cobisim.LAdSAE.invariant}, proceed by structural induction, with the cases for formulas in \LAd (atoms, Boolean operators and $\modDg$) as in Theorem 2.7.  For the remaining case, suppose $(M,w) \cobisim (M',w')$. From left to right, if $(M, w) \Vdash \mmSAE{\sen}{\varphi}$ then, by semantic interpretation, $(\mopSSE{M}{\sen}{\chi}, w) \Vdash \varphi$ holds for every $\chi \in \LAd$. But from $(M,w) \cobisim (M',w')$ and the fact each $\chi$ is in \LAd, it follows that $(\mopSSE{M}{\sen}{\chi},w) \cobisim (\mopSSE{M'}{\sen}{\chi},w')$ for every $\chi \in \LAd$ (\autoref{itm:cobisim.LAdSSE.preserves} in the base case of the proof of Theorem 2.11). Then, by IH, $(\mopSSE{M'}{\sen}{\chi}, w') \Vdash \varphi$ for every $\chi \in \LAd$. Hence, $(M',w') \Vdash \mmSAE{\sen}{\varphi}$. The right-to-left direction is analogous.

For \autoref{itm:cobisim.LAdSAE.preserves}, proceed as in the same case in the proof of Theorem 2.11 , using now the just proved \autoref{itm:cobisim.LAdSAE.invariant} for formulas in $\LAd^\ast$.

\ssparagraph{Inductive case.} As in the same case in the \hyperref[thm:cobisim.LAdSSE:proof]{proof} of \autoref{thm:cobisim.LAdSSE}.

\end{document}